\documentclass[aps,superscriptaddress,twocolumn]{revtex4-1}
\usepackage{amssymb,amsmath,mathptmx}
\usepackage{graphicx}
\usepackage{dcolumn}
\usepackage{bm}
\usepackage{hyperref}
\usepackage{color}
\usepackage[utf8x]{inputenc}
\usepackage{array}
\newcolumntype{C}{>{\centering\arraybackslash}p{1em}}

\newcommand{\be}{\begin{equation}}
\newcommand{\ee}{\end{equation}}
\newcommand{\bea}{\begin{eqnarray}}
\newcommand{\eea}{\end{eqnarray}}
\begin{document}

\title{{Multiterminal ballistic Josephson junctions
    coupled to normal leads}}

\author{R\'egis M\'elin}

\affiliation{Univ. Grenoble-Alpes, CNRS, Grenoble INP\thanks{Institute
    of Engineering Univ. Grenoble Alpes}, Institut NEEL, 38000
  Grenoble, France}

\begin{abstract}
Multiterminal Josephson junctions have aroused considerable
theoretical interest recently and numerous works aim at putting the
predictions of correlations among Coopers (i.e. the so-called
quartets) and simulation of topological matter to the test of
experiments. The present paper is motivated by recent experimental
investigation from the Harvard group reporting $h/4e$-periodic quartet
signal in a four-terminal configuration containing a loop pierced by
magnetic flux, together with inversion controlled by the bias voltage,
i.e. the quartet signal can be larger at half-flux quantum than in
zero magnetic field. Here, we theoretically focus on devices
consisting of finite-size quantum dots connected to four
superconducting and to a normal lead. In addition to presenting
numerical calculations of the quartet signal within a simplified
modeling, we reduce the device to a non-Hermitian Hamiltonian in the
infinite-gap limit. Then, relaxation has the surprising effect of
producing sharp peaks and log-normal variations in the
voltage-sensitivity of the quartet signal in spite of the expected
moderate fluctuations in the two-terminal DC-Josephson current. The
phenomenon is reminiscent of a resonantly driven harmonic oscillator
having amplitude inverse proportional to the damping rate, and of the
thermal noise of a superconducting weak link, inverse proportional to
the Dynes parameter. Perspectives involve quantum bath engineering
multiterminal Josephson junctions in the circuits of cavity quantum
electrodynamics.
\end{abstract}

\maketitle

\section{Introduction}
\label{sec:intro}

BCS superconductors \cite{BCS} are characterized by macroscopic phase
variable $\varphi$ and energy gap $\Delta$ between the ground state
and the first quasiparticles. {Anderson demonstrated
  \cite{Anderson-RPA,Anderson-mass} that Coulomb interaction produces
  collective modes in bulk superconductors, yielding the Higgs
  mechanism as an explanation to the Meissner effect. Josephson
  \cite{Josephson} demonstrated that gauge invariance implies
  DC-supercurrent between two phase-biased superconductors, and
  AC-current oscillations with voltage-biasing. The so-called Andreev
  bound states (ABS) \cite{Andreev,deGennes,SaintJames,Kulik}
  contribute to the DC-Josephson supercurrent flowing through any type
  of weak link, see for instance
  Refs.~\onlinecite{Likharev,AVERIN,Cuevas,Scheer,Beenakker}.
  Recently, the ABS were probed with microwave spectroscopy
  \cite{Bretheau1,Bretheau2,Olivares,Janvier}, but the life-time of
  this ``Andreev qu-bit'' \cite{Zazunov} turns out not to be
  infinite. In general, understanding the mechanisms of quasiparticle
  poisoning
  \cite{Martinis2009,deVisser2011,Lenander2011,Rajauria2012,Wenner2013,Riste2013,LevensonFalk2014,Nazarov-qp}
  is a central issue in the studies of circuits of quantum engineering
  \cite{Kouznetsov,Clarke1,Clarke2,Devoret,Martinis}. In bulk
  superconductors, the electron-phonon coupling or the
  electron-electron repulsive Coulomb interaction produce finite
  quasiparticle life-time \cite{Kaplan,Dynes,Pekola1,Pekola2} captured
  by adding small imaginary part $\eta$ to their energy. This small
  Dynes parameter $\eta$ \cite{Kaplan,Dynes,Pekola1,Pekola2} produces
  exponential time-decay of the quasiparticle wave-functions. In the
  following paper, we investigate multiterminal Josephson junctions
  coupled to normal leads, and demonstrate that small relaxation
  produced by the coupling to those normal conductors has drastic
  effect on the current. Based on previous works related to
  infinite-gap Hamiltonians \cite{Zazunov,Meng,topo1_plus_Floquet}, we
  also here-consider the infinite-gap limit, where non-Hermitian
  Hamiltonians emerge for quantum dots coupled to superconducting and
  normal leads. Those non-Hermitian Hamiltonians receive
  interpretation of describing the coherent oscillations and damping
  in the ABS dynamics, and they capture the degrees of freedom of the
  ABS having spatial extent set by the BCS coherence length, which is
  vanishingly small in the infinite-gap limit. The present paper also
  suggests the interest of future quantum bath engineering for
  multiterminal Josephson junctions connected to cavity-quantum
  electrodynamics resonators.}

{Now, we specifically introduce the considered
  four-terminal Josephson junctions, starting with the three-terminal
  Cooper pair beam splitters that have been the subject of intense
  investigations since more than twenty years. Entangled Andreev pairs
  can split if two independent voltages are applied in three-terminal
  $F_aSF_b$ ferromagnet-superconductor-ferromagnet or $N_aSN_b$ normal
  metal-superconductor-normal metal Cooper pair beam splitters
  \cite{theory-CPBS1,theory-CPBS2,theory-CPBS3,theory-CPBS4,theory-CPBS4-bis,theory-CPBS4-ter,theory-CPBS5,theory-CPBS6,theory-CPBS7,theory-CPBS8,theory-CPBS9,theory-CPBS10,theory-CPBS11,Yeyati2007,Danneau,Beckmann2007,Schindele2012,Schindele2014,Tan2015,Borzenets2016a,exp-CPBS1,exp-CPBS2,exp-CPBS3,exp-CPBS4,exp-CPBS5,exp-CPBS6,exp-CPBS7,exp-CPBS8}. ``Nonlocal''
  or ``crossed'' Andreev reflection (CAR) appears if the separation
  between the $N_aS$ and $SN_b$ interfaces is comparable to the
  zero-energy superconducting coherence length. Spin-up electron from
  lead $N_b$ can be Andreev reflected as spin-down hole into $N_a$,
  leaving a Cooper pair in the ``central'' $S$. This yields nonlocal
  current response reflecting production of nonlocally split
  Cooper pairs. In addition, the positive current-current
  cross-correlations of Cooper pair splitting
  \cite{theory-CPBS7,theory-noise1,theory-noise2,theory-noise3,theory-noise4,theory-noise5,theory-noise6,theory-noise7,theory-noise8,theory-noise9,theory-noise10,theory-noise11,theory-noise12}
  were experimentally revealed \cite{exp-CPBS7,exp-CPBS8}.}

{The double quantum dot experiments
  \cite{exp-CPBS8,exp-CPBS6,exp-CPBS5} provide evidence for nonlocal
  two-particle Andreev resonance in $N_a$-dot-$S_c$-dot-$N_b$ devices,
  on the condition of the opposite energy levels
  $\epsilon_a=-\epsilon_b$ on both quantum dots.} In the following, we
demonstrate how ``nonlocal two-particle Andreev resonance'' in
$N_a$-dot-$S_c$-dot-$N_b$ Cooper pair beam splitters can be
generalized to ``nonlocal two-Cooper pair resonance'' in
all-superconducting three-terminal Josephson junctions.

Those two-Cooper pair resonances build on nonlocal two-Cooper pair
states, the so-called quartets
\cite{Cuevas-Pothier,Freyn,Melin1,Jonckheere,FWS,Sotto,engineering,paperI,paperII,long-distance,Akh2,Pillet,Pillet2,Nazarov-PRR,Nazarov-PRB-AM}
that appear in $(S_a,\,S_c,\,S_b)$ three-terminal Josephson junctions,
where $S_a$ and $S_b$ are voltage-biased at $V_a$ and $V_b$, the
superconducting lead $S_c$ being grounded at $V_c=0$. Andreev
scattering yields quartet phase-sensitive DC-current response if the
condition $V_a=-V_b\equiv V$ is fulfilled. On this quartet line
$V_a=-V_b$, the elementary transport process transfers two Cooper
pairs from $S_a$ and $S_b$ into the grounded $S_c$, while exchanging
partners
\cite{Cuevas-Pothier,Freyn,Melin1,Jonckheere,FWS,Sotto,engineering,paperI,paperII,long-distance,Akh2,Pillet,Pillet2,Nazarov-PRR,Nazarov-PRB-AM}. By
the time-energy uncertainty relation, this phase-sensitive quartet
current is DC, since, on the quartet line, the energy
$E_i=2e(V_a+V_b)$ of two Cooper pairs from $S_a$ and $S_b$ in the
initial state is equal to the final state energy $E_f=4eV_c$ of the
two Cooper pairs transmitted into $S_c$.

\begin{figure}[htb]
  \includegraphics[width=.5\columnwidth]{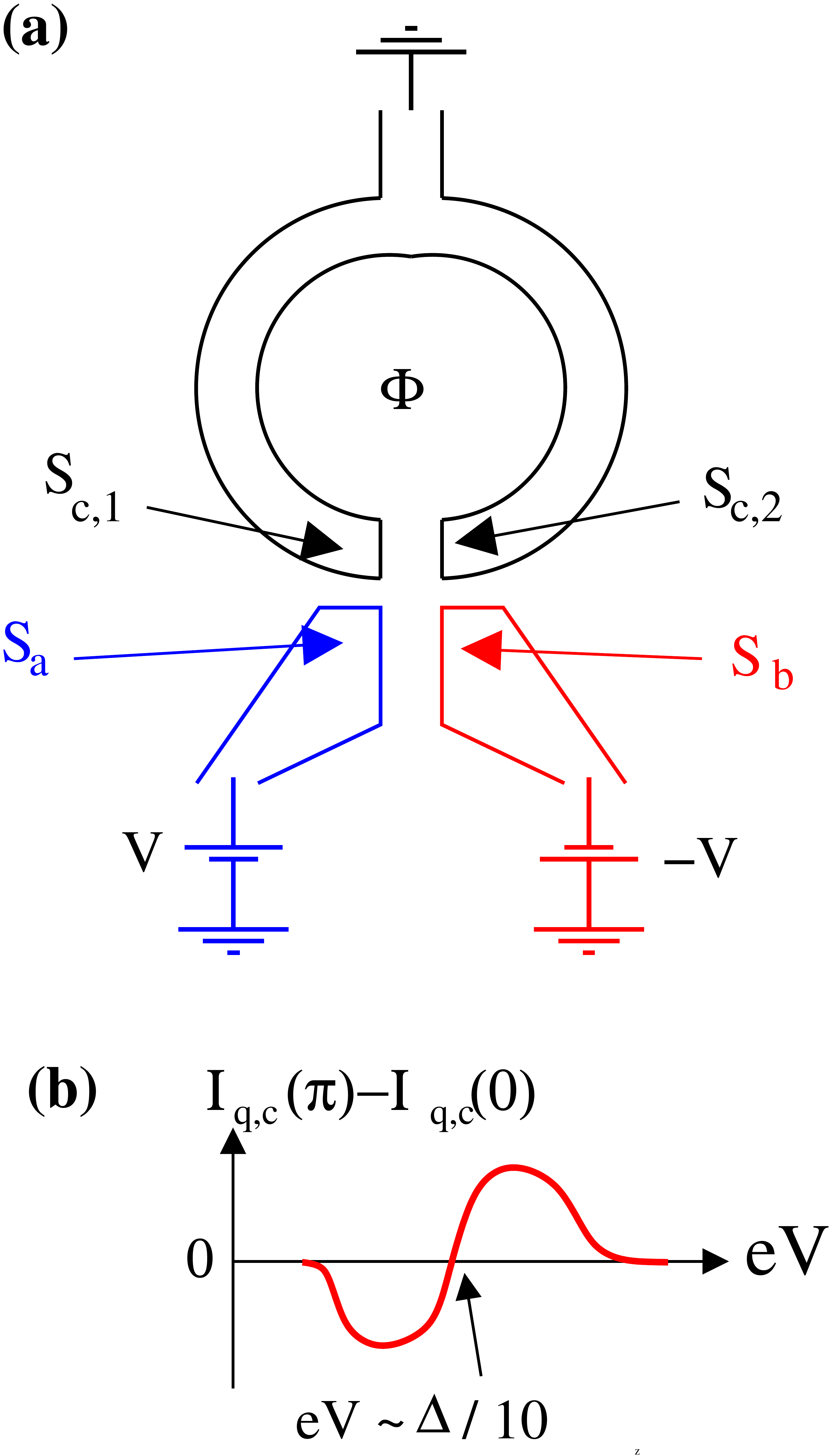}
  \caption{{\it Schematic four-terminal experiment:} Panel a
    represents top-view of the experimental device, with the
    superconducting leads $S_a$ and $S_b$ biased at $V_{a,b}=\pm
    V$. The loop terminated by $S_{c,1}$ and $S_{c,2}$ is connected to
    ground by the contact on top of it, and it is pierced by the
    magnetic field flux $\Phi$. The four superconducting leads are
    evaporated on a graphene layer gated far from the Dirac point, and
    forming a two-dimensional (2D) metal. Panel b summarizes the
    experimental result \cite{HGE} that inspires the present
    theoretical paper, i.e. the quartet critical current current
    $I_{q,c}(\pi)$ at flux $\Phi=\pi$ can be larger or smaller than
    $I_{q,c}(0)$ at flux $\Phi=0$, in a way that is sensitive to
    voltage typically being in the range $eV\approx \Delta/10$ where
    $\Delta$ is the superconducting gap.
  \label{fig:experiment}
  }
\end{figure}

Several experiments were recently performed on multiterminal Josephson
junctions
\cite{Lefloch,Heiblum,HGE,multiterminal-exp1,multiterminal-exp2,multiterminal-exp3,multiterminal-exp4,multiterminal-exp5,multiterminal-exp6,multiterminal-exp7,multiterminal-exp8,multiterminal-exp9}.
Some of those explored the possibility of nontrivial topology
\cite{vanHeck,Padurariu,multiterminal-exp1,Nazarov1,Nazarov2,topo0,topo1,topo2,topo3,topo4,topo4-bis,topo5,Berry,Feinberg1,Feinberg2,topo1_plus_Floquet,Levchenko1,Levchenko2,Akh2}. On
the other hand, three groups reported compatibility with the quartets
\cite{Cuevas-Pothier,Freyn,Melin1,Jonckheere,FWS,Sotto,engineering,paperI,paperII,long-distance,Akh2,Pillet,Pillet2,Nazarov-PRR,Nazarov-PRB-AM}:
the Grenoble group experiment with metallic structures \cite{Lefloch},
the Weizmann Institute group experiment with semiconducting nanowires
\cite{Heiblum} and the more recent Harvard group experiment on
ballistic graphene-based four-terminal Josephson junctions \cite{HGE}.
{This third experiment \cite{HGE} is summarized in
  figures~\ref{fig:experiment}a and~\ref{fig:experiment}b.  The latter
  shows typical experimental variations for the quartet critical
  current as a function of the bias voltage $V$ and the magnetic flux
  $\Phi$ through the loop.  The experiment features the
  counterintuitive voltage-$V$-tunable inversion, i.e. the possibility
  of stronger quartet critical current at half-flux quantum $\Phi=\pi$
  than in zero field $\Phi=0$, see figure~\ref{fig:experiment}b. The
  present paper is motivated by exploring interpretation of this
  experiment, in the continuation of our previous papers~I and~II
  \cite{paperI,paperII}.} {We also underline two
  recent experimental preprints reporting the quartet line in
  ballistic devices \cite{multiterminal-exp8,multiterminal-exp9}.}

{Zero-dimensional (0D) quantum dots were used in
  Ref.~\onlinecite{Park} to model two-terminal graphene-superconductor
  hybrids, in connection with experimental evidence for the Floquet
  replica of the Andreev spectrum under microwave radiation. The goal
  of the present paper is to similarly describe four-terminal
  graphene-based Josephson junctions with effective models based on
  quantum dots{, in the absence of Coulomb
    interaction}. We argue that multilevel quantum dots coupled to
  four superconducting and to a normal lead are ``minimal models'' of
  four-terminal Josephson junctions that are intermediate between the
  short- and long-junction limits.} In addition, we further simplify
the description in terms of double 0D quantum dots. Four or more ABS
are within the gap in the equilibrium limit, instead of two ABS for a
single 0D quantum dot.

{Importantly, nonproximitized regions can appear in the
  two-dimensional (2D) metal, typically in the ``cross-like region''
  formed in between the contacts with the superconductors, see
  figure~\ref{fig:experiment}a. Then, the quantum transport process
  can have initial or final states in those normal regions of the
  circuit. Current lines can be converted by Andreev reflection as
  supercurrent flowing in the grounded superconducting loop. Namely,
  current conservation for symmetric coupling to the superconducting
  leads implies that the 2D metal chemical potential is at the same
  zero energy as in the superconducting loop. Thus, the normal carriers
  transmitted through the 2D metal can be transferred into the
  grounded superconducting loop without taking or giving energy.}

{Now, we explain the connection between the present work and our
  previous papers.}  Compared to our previous Ref.~\onlinecite{FWS},
{we implement} a double 0D quantum dot, instead of the single quantum
dot Dynes parameter \cite{Kaplan,Dynes,Pekola1,Pekola2} {model of}
Ref.~\onlinecite{FWS}. We also include four superconducting leads,
instead of three in our previous Ref.~\onlinecite{FWS}. {Our previous
  paper~I \cite{paperI} presented an expansion of the quartet critical
  current in perturbation in the tunneling amplitudes between a 2D
  metal and four superconducting leads. The present paper addresses
  resonances that appear beyond the weak coupling limit.} {Our}
previous paper~II \cite{paperII} treated single quantum dot with
relaxation solely originating from the continua of BCS
quasiparticles. Relaxation then produces nonvanishingly small
line-width broadening {$\delta_0$} for the Floquet resonances, which
behaves like {$\log \delta_0 \sim -\Delta/eV$} at low bias voltage
energy $eV$ compared to the superconducting gap $\Delta$
\cite{FWS,engineering,Berry,paperII}. In the four-terminal device of
paper~II \cite{paperII}, Landau-Zener quantum tunneling reduces the
quartet current by coherent superpositions in the dynamics of the two
opposite current-carrying ABS, at voltage values that are close to
avoided crossings in the Floquet spectra. This mechanism is analogous
to the reduction of the DC-Josephson current in microwave-irradiated
weak links \cite{Bergeret}.  Conversely, in the present work, the
interplay between the time-periodic dynamics and finite line-width
broadening due to the attached normal lead {restores the expected}
sharp resonance peaks in the voltage-dependence of the quartet
current. For instance, a resonantly driven harmonic oscillator has
amplitude inverse proportional to the damping rate. In addition, we
note that the zero-frequency noise of a single superconducting weak
link at finite temperature is inverse proportional to the rate set by
the Dynes parameter \cite{Kaplan,Dynes,Pekola1,Pekola2} in some
parameter window, see Ref.~\onlinecite{Yeyati-noise-equilibrium}. The
same scaling holds in this reference for the noise at frequency
$2\omega_S$, where $\omega_S$ is the ABS energy. The quartets in
four-terminal devices \cite{HGE,paperI,paperII} offer several
parameters to probe this physics, such as the quartet phase variable
and the ``knobs'' of the bias voltage and magnetic flux.

The paper is organized as follows. The Hamiltonians are presented in
section~\ref{sec:H}.  The mechanism for the two-Cooper pair resonance
is discussed in section~\ref{sec:mechanism}.  The mechanism for the
inversion is presented in section~\ref{sec:mecha-inversion}.
Section~\ref{sec:num-2nodes} shows numerical results. The low-voltage
limit is discussed in section~\ref{sec:low-V}. Concluding remarks are
presented in the final section~\ref{sec:conclusions}.

\section{Hamiltonians}
\label{sec:H}
{In this section, we provide the Hamiltonians on which
  the paper is based. Subsection~\ref{sec:Hamiltonians} presents the
  full Hamiltonians of the superconductors, quantum dots and normal
  leads. It goes beyond the goal of the present paper to produce
  numerical results for the voltage-biased four-terminal Josephson
  junctions described by the general Hamiltonians of
  subsection~\ref{sec:Hamiltonians} in a realistic geometry. This is
  why we present in subsection~\ref{sec:model-H} simplified model
  Hamiltonians which will subsequently be used for numerical
  calculations.}

\subsection{General Hamiltonians}
\label{sec:Hamiltonians}

{Now, we introduce in subsection~\ref{sec:H-supra} the
  BCS Hamiltonian of the superconducting leads.}
Subsections~\ref{sec:H-qu-dots} and~\ref{sec:H-normal} introduce the
Hamiltonians of the quantum dots and normal leads
respectively. {As it is mentioned above, the numerical
  calculations presented below are based on simplifications of those
  Hamiltonians.}

\subsubsection{Superconductors}
\label{sec:H-supra}

{The Hamiltonian of a BCS superconductor is the
  following:}
\begin{eqnarray}
  \label{eq:H-BCS1}
        {\cal H}_{BCS}&=&-W \sum_{\langle i,j \rangle}
        \sum_{\sigma_z=\uparrow,\downarrow} \left(c_{i,\sigma_z}^+
        c_{j,\sigma_z}+ c_{j,\sigma_z}^+ c_{i,\sigma_z}\right) \\&-&
        \Delta \sum_i \left(\exp\left(i\varphi_i\right)
        c_{i,\uparrow}^+ c_{i,\downarrow}^+ +
        \exp\left(-i\varphi_i\right) c_{i,\downarrow}
        c_{i,\uparrow}\right) ,
        \nonumber
\end{eqnarray}
where $\sum_{\langle i,j \rangle}$ denotes summation over pairs of
neighboring tight-binding sites labeled by $i$ and $j$, and $\sigma_z$
is the component of the spin along quantization axis. The bulk hopping
amplitude is denoted by $W$ and the superconducting gap $\Delta$ is
taken identical in all superconducting leads $S_a$, $S_b$, $S_{c,1}$
and $S_{c,2}$. The variable $\varphi_i$ denotes the superconducting
phase variable at the tight-binding site labeled by $i$. In the
following, the current is weak and $\varphi_i$ is approximated as
being uniform in space, with the values $\varphi_a$, $\varphi_b$,
$\varphi_{c,1}$, and $\varphi_{c,2}$ in $S_a$, $S_b$, $S_{c,1}$ and
$S_{c,2}$ respectively.

We assume short distance between the contact points $c_1$ and $c_2$
(at the interfaces between $S_{c,1}$ or $S_{c,2}$ and the quantum
dot), and we use the approximation of the gauge
\begin{eqnarray}
  \label{eq:gauge1}
\varphi_{c,1}&=&\varphi_{c}\\
\varphi_{c,2}&=&\varphi_{c}+\Phi
,
\label{eq:gauge2}
\end{eqnarray}
where $\Phi$ is the flux enclosed in the loop terminated by $S_{c,1}$
and $S_{c,2}$.

\subsubsection{Double quantum dots}
\label{sec:H-qu-dots}

{A legitimate approximation is to discard the Coulomb
  electron-electron repulsion at high transparency. Then the
  finite-size quantum dot is described by the Hamiltonian
\begin{eqnarray}
  \label{eq:qu-dot-H}
        {\cal H}_{dot}&=&-W \sum_{\langle i,j \rangle}
        \sum_{\sigma_z=\uparrow,\downarrow} \left(c_{i,\sigma_z}^+
        c_{j,\sigma_z}+ c_{j,\sigma_z}^+ c_{i,\sigma_z}\right) ,\\
        &&- \epsilon_g \sum_{i}
        \sum_{\sigma_z=\uparrow,\downarrow} c_{i,\sigma_z}^+
        c_{i,\sigma_z}
\end{eqnarray}
where the hopping amplitude $W$ is identical to Eq.~(\ref{eq:H-BCS1}),
and $\epsilon_g$ is proportional to the gate voltage. The summation
over pairs of neighboring tight-binding sites is restricted to the
region of finite dimension defining the quantum dot.}

{Conversely, the Hamiltonian of a single 0D quantum
  dot $D_x$ at location ${\bf x}$ is the following:}
\begin{equation}
  \label{eq:H-x-0D}
  {\cal H}_{D_x, 0D}=\epsilon_x \sum_{\sigma_z} c_{D_x,\sigma_z}^+ c_{D_x,\sigma_z}
  ,
\end{equation}
where $\epsilon_x$ is the on-site energy. Similarly, the double
quantum dot $(D_x,D_y)$ is characterized by the on-site energies
$\epsilon_x$ and $\epsilon_y$ with tunneling amplitude $\Sigma^{(0)}$
between them:
\begin{eqnarray}
  \label{eq:H-y-0D}
  {\cal H}_{D_y, 0D}&=&\epsilon_y \sum_{\sigma_z} c_{D_y,\sigma_z}^+
  c_{D_y,\sigma_z}\\ {\cal H}_{T,0}&=&-\Sigma^{(0)} \sum_{\sigma_z}
  c_{D_x,\sigma_z}^+ c_{D_y,\sigma_z} + h.c.
  \label{eq:H-xy}
\end{eqnarray}

At the exception of one of the numerical calculations presented in
section~\ref{sec:num-2nodes}, we use $\epsilon_x = \epsilon_y=0$ in
the paper, thus with
\begin{eqnarray}
  \label{eq:H-dot-1}
  {\cal H}_{D_x, 0D} &=& 0\\
  {\cal H}_{D_y, 0D}&=&0.
\label{eq:H-dot-2}
\end{eqnarray}

\subsubsection{Normal leads}
\label{sec:H-normal}

{The Hamiltonian of a normal lead is given by
\begin{eqnarray}
  \label{eq:H-N}
        {\cal H}_{N}&=&-W \sum_{\langle i,j \rangle}
        \sum_{\sigma_z=\uparrow,\downarrow} \left(c_{i,\sigma_z}^+
        c_{j,\sigma_z}+ c_{j,\sigma_z}^+ c_{i,\sigma_z}\right)
        \nonumber
\end{eqnarray}
where the normal lead chemical potential is vanishingly small, and the
hopping amplitude $W$ is identical to
Eq.~(\ref{eq:H-BCS1}). Eq.~(\ref{eq:H-N}) is similar to the quantum
dot Hamiltonian given by Eq.~(\ref{eq:qu-dot-H}) but now, the
summation over the pairs of neighboring tight-binding sites runs over
an infinite lattice.}

\begin{figure*}[htb]
  \begin{minipage}{.7\textwidth}
    \includegraphics[width=.75\textwidth]{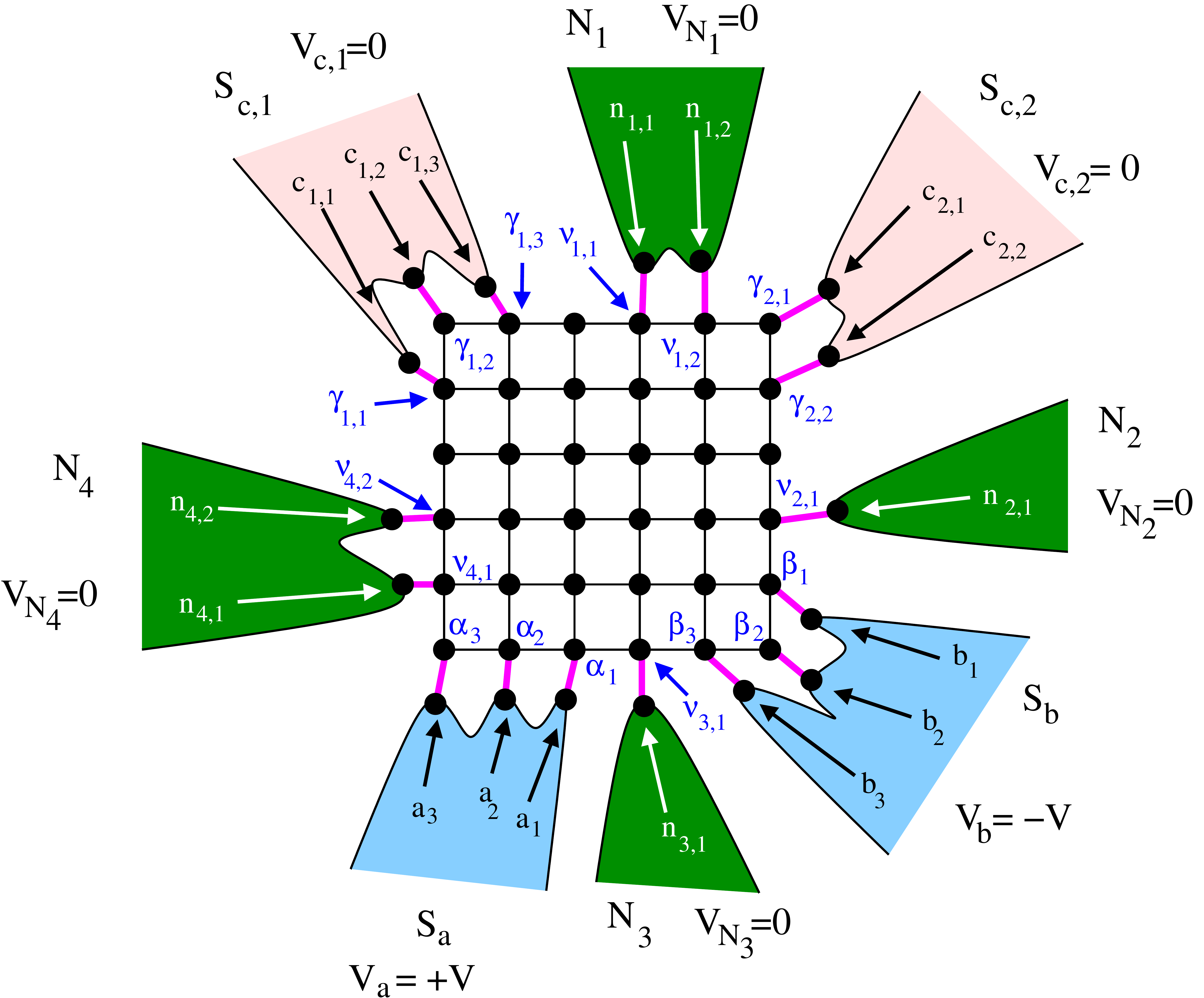}
    \end{minipage}\begin{minipage}{.28\textwidth}
  \caption{{\it The four-terminal superconducting multilevel quantum
      dot:} The figure shows a multilevel quantum dot connected to the
    four superconducting leads $S_a$, $S_b$, $S_{c,1}$ and $S_{c,2}$
    biased at $V_{a,b}=\pm V$ and $V_{c,1}=V_{c,2}=0$, and to the four
    grounded normal leads $N_1$, $N_2$, $N_3$ and $N_4$. The tunneling
    between the multilevel quantum dot and the superconducting leads
    is provided in
    Eqs.~(\ref{eq:tunneling-S-1})-(\ref{eq:Tmulti-c2}). The tunneling
    to the normal leads is provided in
    Eq.~(\ref{eq:tunneling-N}). This model will be used in the
    following sections~\ref{sec:2CPR}, \ref{sec:closing-Dyson},
    \ref{sub2}, \ref{sec:transport} and
    \ref{sec:current-conservation}.
  \label{fig:geometrie-4T-multilevel}
  }\end{minipage}
\end{figure*}

\begin{figure}[htb]
  \includegraphics[width=.75\columnwidth]{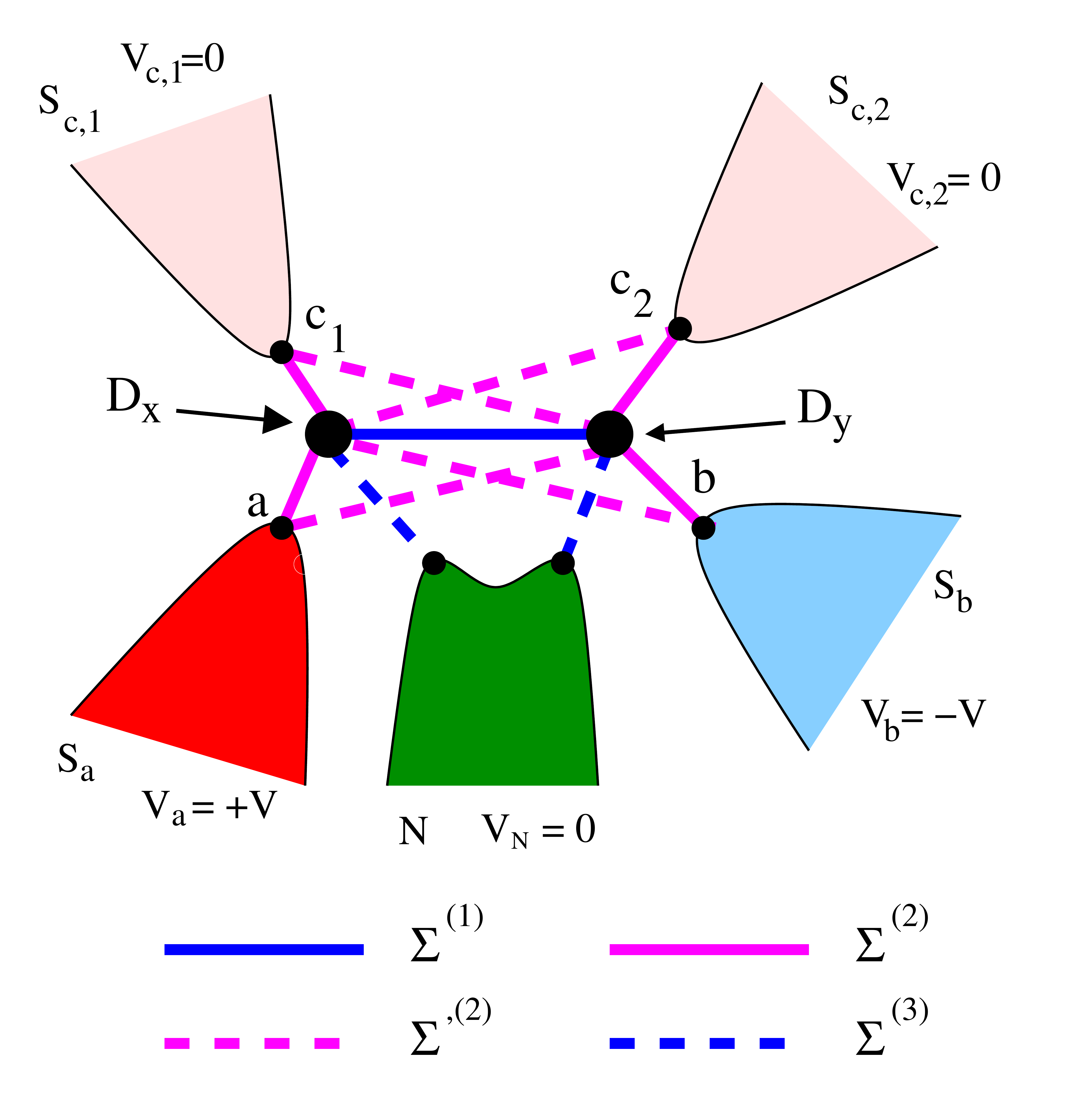}
  \caption{{\it The four-terminal phenomenological model of
      superconducting double 0D quantum dot with normal lead in
      parallel:} The figure shows a double 0D quantum dot connected to
    the four superconducting leads $S_a$, $S_b$, $S_{c,1}$ and
    $S_{c,2}$ biased at $V_{a,b}=\pm V$ and $V_{c,1}=V_{c,2}=0$, and
    to the four grounded normal leads $N_1$, $N_2$, $N_3$ and
    $N_4$. The corresponding Hamiltonian is provided in
    section~\ref{sec:closely-spaced-double}. The coupling
    $\Sigma^{(0)}$ directly couples the two quantum dots $D_x$ and
    $D_y$, see Eq.~(\ref{eq:H-xy}). The ``direct'' coupling
    $\Sigma^{(2)}$ connects $D_x$ to $S_a$, $D_x$ to $S_{c,1}$, $D_y$
    to $S_b$ and $D_y$ to $S_{c,2}$, see
    Eqs.~(\ref{eq:Tclosely-spaced-a})-(\ref{eq:Sigma(2)}).The
    ``crossed'' coupling $\Sigma'^{(2)}$ connects $D_x$ to $S_b$,
    $D_x$ to $S_{c,2}$, $D_y$ to $S_a$ and $D_y$ to $S_{c,1}$, see
    Eqs.~(\ref{eq:Tclosely-spaced-b-crossed})-(\ref{eq:Sigma-prime-2}).
    The coupling $\Sigma^{(3)}$ connects the 0D quantum dot $D_x$ to
    $N_x$ and $D_y$ to $N_y$, where $N_x$ and $N_y$ are the two
    tight-binding sites on the normal lead $N$-side of the contacts,
    see Eqs.~(\ref{eq:coupling-N1})-(\ref{eq:Sigma-3}). This model
    will be used in the following sections~\ref{sec:2CPR},
    \ref{sec:closing-Dyson}, \ref{sub2}, \ref{sec:specializing},
    \ref{sec:transport} and \ref{sec:current-conservation}.
  \label{fig:geometrie-4T-double-dot}
  }
\end{figure}

\subsection{Simplified model Hamiltonians}

\label{sec:model-H}

{Our previous paper~I \cite{paperI} was based on}
perturbation theory in the tunnel amplitudes between a 2D metal and
four superconducting leads, in the $V=0^+$ adiabatic limit. In this
paper~I, the poles of the Green's functions are at the gap edge
singularities and those perturbative calculations do not yield small
voltage scales in the quartet critical current-voltage
characteristics. In our previous paper~II, we addressed the connection
between the quartet critical current and the Floquet spectra in a
situation where 0D quantum dot is connected to four superconducting
leads.  We then obtained reduction of the quartet critical current at
the bias voltage resonance values. Compared to this paper~II, the
here-considered four-terminal double quantum dot Josephson junction is
additionally connected to normal lead.

{In this section, we provide the Hamiltonians of the three considered
simplified models, the third one being numerically implemented in the
forthcoming section~\ref{sec:num-2nodes}.}

Subsection~\ref{sec:multilevel-quantum-dot} deals with a multilevel
quantum dot connected to four superconducting and to normal
leads. Subsections~\ref{sec:closely-spaced-double}
and~\ref{sec:remote-0D} deal with two simple models of double 0D
quantum dots, also attached to superconducting and normal
leads. Subsections~\ref{sec:closely-spaced-double}
and~\ref{sec:remote-0D} correspond to ``double 0D quantum dot
connected to normal lead'', in parallel or in series respectively. The
interest of double 0D quantum dots with respect to single, triple or
quadruple 0D quantum dots is explained in
subsection~\ref{sec:comments-choice}.

\subsubsection{Multilevel quantum dot}
\label{sec:multilevel-quantum-dot}

{Now, we consider the Hamiltonian of a multilevel
  quantum dot connected to superconducting and normal leads. We also
  provide a physical discussion. This defines a first stage in
  reducing the four-terminal device to simpler Hamiltonians.}

We assume that the multilevel quantum dot on
figure~\ref{fig:geometrie-4T-multilevel} is intermediate between the
short- and long-junction limits, i.e. it has dimension $\agt 2
\xi_0$, where $\xi_0=\hbar v_F/\Delta$ is the BCS coherence length,
with $v_F$ the Fermi velocity. Thus, more than two ABS are formed at
equilibrium.

{\it The Hamiltonian:} {We consider finite-size discrete levels on the
  multilevel quantum dot of figure~\ref{fig:geometrie-4T-multilevel},
  at the energies $\Omega_\psi$. The corresponding Hamiltonian takes
  the form
\begin{equation}
  \label{eq:H-multi}
  {\cal H}_{multi}=\sum_\psi \Omega_\psi \sum_{\sigma_z}
  c_{\psi,\sigma_z}^+ c_{\psi,\sigma_z} ,
\end{equation}
where $c_{\psi,\sigma_z}^+$ creates a fermion with spin-$\sigma_z$ in
the state $|\psi\rangle$. The superconducting leads $S_a$, $S_b$,
$S_{c,1}$ and $S_{c,2}$ are described by the BCS Hamiltonian given by
Eq.~(\ref{eq:H-BCS1}). The normal lead is described by
Eq.~(\ref{eq:H-N}). Tunneling between the leads $(S_a, S_b,
S_{c,1},S_{c,2})$ biased at $(V,-V,0,0)$ and the multilevel quantum
dot is described by multi-channel contacts at the time~$\tau$:
\begin{eqnarray}
  \label{eq:tunneling-S-1}
        {\cal H}_{T,a}(\tau)&=& -\Sigma^{(1)}_a \exp\left(-i eV\tau/\hbar\right)
        \sum_s \sum_{\sigma_z} c_{a_s,\sigma_z}^+
  c_{\alpha_s,\sigma_z} \\\nonumber&&+ h.c.\\
  {\cal H}_{T,b}(\tau)&=& -\Sigma^{(1)}_b \exp\left(i eV\tau/\hbar\right)
  \sum_t \sum_{\sigma_z} c_{b_t,\sigma_z}^+
  c_{\beta_t,\sigma_z} \\\nonumber&&+ h.c.\\
{\cal H}_{T,c_1}&=& -\Sigma^{(1)}_{c_1} \sum_u \sum_{\sigma_z} c_{{c_{1,u}},\sigma_z}^+
c_{\gamma_{1,u},\sigma_z} + h.c.\\
{\cal H}_{T,c_2}&=& -\Sigma^{(1)}_{c_2} \sum_v \sum_{\sigma_z} c_{{c_{2,v}},\sigma_z}^+
c_{\gamma_{2,v},\sigma_z} + h.c.
\label{eq:Tmulti-c2},
\end{eqnarray}
where the integers $s,\,t,\,u$ and $v$ label the collection of hopping
amplitudes in real space, thus realizing a multichannel interface. As
an approximation, the hopping amplitudes in
Eqs.~(\ref{eq:tunneling-S-1})-(\ref{eq:Tmulti-c2}) were chosen
identical within each contact: $\Sigma^{(1)}_{a,s}\equiv
\Sigma^{(1)}_a$, $\Sigma^{(1)}_{b,t}\equiv \Sigma^{(1)}_b$,
$\Sigma^{(1)}_{c_1,u}\equiv \Sigma^{(1)}_{c_1}$ and
$\Sigma^{(1)}_{c_2,v}\equiv \Sigma^{(1)}_{c_2}$. The contact
transparencies are parameterized by}
{
  \begin{eqnarray}
    \label{eq:Gamma-a}
\Gamma_a&=&\left(\Sigma^{(1)}_a\right)^2/W,\\
\Gamma_b&=&\left(\Sigma^{(1)}_b\right)^2/W,\\
\Gamma_{c_1}&=&\left(\Sigma^{(1)}_{c_1}\right)^2/W,\\
\Gamma_{c_2}&=&\left(\Sigma^{(1)}_{c_2}\right)^2/W.
\label{eq:Gamma-c2}
\end{eqnarray}}
{The notations
  $c_{a_s,\sigma_z}^+$, $c_{b_t,\sigma_z}^+$,
  $c_{{c_{1,u}},\sigma_z}^+$ and $c_{{c_{2,v}},\sigma_z}^+$ refer to
  creating a spin-$\sigma_z$ fermion at the tight-binding sites
  labeled by $a_s$, $b_t$, $c_{1,u}$ and $c_{2,v}$ belonging to the
  superconducting leads $S_a$, $S_b$, $S_{c,1}$ and $S_{c,2}$, see
  $(a_1,a_2,a_3)$, $(b_1,b_2)$, $(c_{1,1},c_{1,2},c_{1,3})$ and
  $(c_{2,1},c_{2,2})$ in
  figure~\ref{fig:geometrie-4T-multilevel}. Similarly,
  $c_{\alpha_s,\sigma_z}$, $c_{\beta_t,\sigma_z}$,
  $c_{\gamma_{1,u},\sigma_z}$ and $c_{\gamma_{2,v},\sigma_z}$ destroy
  a spin-$\sigma_z$ fermion at the tight-binding sites labeled by
  $\alpha_s$, $\beta_t$, $\gamma_{1,u}$ and $\gamma_{2,v}$, see
  $(\alpha_1,\alpha_2,\alpha_3)$, $(\beta_1,\beta_2)$,
  $(\gamma_{1,1},\gamma_{1,2},\gamma_{1,3})$ and
  $(\gamma_{2,1},\gamma_{2,2})$ in
  figure~\ref{fig:geometrie-4T-multilevel}.}

{We assume in addition that $q_0$ normal leads labeled by
  $n_q=1,...,q_0$ are connected to the multilevel quantum dot by
  the following tunneling Hamiltonian:
  \begin{equation}
    \label{eq:tunneling-N}
  {\cal H}_{T,N_q}= -\Sigma^{(1)}_{N_q} \sum_{w} \sum_{\sigma_z} c_{n_{q,w},\sigma_z}^+
  c_{\nu_{q,w},\sigma_z} + h.c.
  ,
\end{equation}
where $n_{q,w}$ is the tight-binding site labeled by $w$ in the lead
$N_q$, and $\nu_{q,w}$ is its counterpart on the multilevel quantum
dot. We took in Eq.~(\ref{eq:tunneling-N}) the same tunneling
amplitudes $\Sigma^{(1)}_{N_q}\equiv \Sigma^{(1)}_{N_q,w}$ at each
contact. For instance, for $q_0=4$, the tight-binding sites
$(n_{1,1},n_{1,2})$, $n_{2,1}$, $n_{3,1}$ and $(n_{4,1}, n_{4,2})$
belonging to the normal leads $N_1$, $N_2$, $N_3$ and $N_4$
respectively are shown in figure~\ref{fig:geometrie-4T-multilevel},
together with their counterparts $(\nu_{1,1},\nu_{1,2})$, $\nu_{2,1}$,
$\nu_{3,1}$ and $(\nu_{4,1}, \nu_{4,2})$ in the multilevel quantum
dot.}

{This model will be used in the following sections~\ref{sec:2CPR},
  \ref{sec:closing-Dyson}, \ref{sub2}, \ref{sec:transport} and
  \ref{sec:current-conservation}.}

{\it Physical remarks:} Two antagonist effects appear:

(i) In absence of coupling to the superconductors, increasing the
coupling to the normal leads has the tendency to smoothen the
energy-dependence of the local density of states.

(ii) In the presence of coupling to the superconducting leads, the
Floquet resonances have the tendency to produce sharp peaks in the local
density of states as a function of energy.

Physically, the above item (i) captures the metallic limit of ``weak
sample-to-sample fluctuations in the quartet critical current''. The
item (ii) corresponds to ``strong sample-to-sample fluctuations in the
quartet current'', where samples differ by the number of channels at
the contacts, or by fluctuations in the shape of the multilevel
quantum dot. The metallic limit (i) is ruled out for compatibility
with the experiment in Ref.~\onlinecite{HGE}, because it does not
produce small voltage scales. This is why we focus here on the regime
(ii) of weak coupling to the normal leads.

{In most of the work, we make an additional assumption
  about the spectrum $\{\Omega_\psi\}$ of the multilevel quantum dot,
  see Eq.~(\ref{eq:H-multi}). Namely, we assume pairs of levels at
  opposite energies, i.e.  there are values $\psi_1$, $\psi_2$ of the
  label $\psi$ such that $\Omega_{\psi_1}=-\Omega_{\psi_2}$. Then,
  two-Cooper pair resonance emerges at specific values of the bias
  voltage, such as $2 eV_*= \Omega_{\psi_1}-\Omega_{\psi_2} = 2
  \Omega_{\psi_1}$ in the limit of weak coupling. However, within the
  considered models, we will find in section~\ref{sec:num-2nodes}
  robustness of the sharp resonances with respect to detuning from the
  condition of opposite energy levels.}
  
{Next, we simplify} the multilevel quantum dot into
phenomenological models of double 0D quantum dots that seem to be the
``minimal models'' for the two-Cooper pair resonance.

\subsubsection{Phenomenological model of double 0D
  quantum dot with normal lead in parallel}
\label{sec:closely-spaced-double}

Now, we present the Hamiltonian of a double 0D quantum dot with normal
lead in parallel, and additionally connected to superconducting
leads. We also provide a physical discussion. {This
  defines a second stage in proposing models that can practically be
  numerically implemented.}

{{\it The Hamiltonian:} {Specifically, we consider a
    second model,} i.e. a ``phenomenological model of double 0D
  quantum dot with normal lead in parallel'', see
  figure~\ref{fig:geometrie-4T-double-dot}. The quantum dots $D_x$ and
  $D_y$ have the vanishingly small Hamiltonians of
  Eqs.~(\ref{eq:H-dot-1}) and~(\ref {eq:H-dot-2}). The coupling
  between $D_x$ and $D_y$ is given by Eq.~(\ref{eq:H-xy}) with the
  hopping amplitude $\Sigma^{(0)}$ shown as a light blue line in
  figure~\ref{fig:geometrie-4T-double-dot}. Again, the superconducting
  leads $S_a$, $S_b$, $S_{c,1}$ and $S_{c,2}$ are described by the BCS
  Hamiltonian, see Eq.~(\ref{eq:H-BCS1}).  Tunneling between the leads
  $S_a$, $S_b$, $S_{c,1}$ and $S_{c,2}$ and the double 0D quantum dot
  $(D_x,D_y)$ is described by single-channel contacts:
\begin{eqnarray}
 \label{eq:Tclosely-spaced-a}
 {\cal H}_{T,D_x,a}(\tau)&=& -\Sigma^{(2)}_{D_x,a} \exp\left(-i eV\tau/\hbar\right) \sum_{\sigma_z}
  c_{a,\sigma_z}^+ c_{D_x,\sigma_z} \\\nonumber&&+ h.c.\\
  {\cal H}_{T,D_y,b}(\tau)&=&
  -\Sigma^{(2)}_{D_y,b} \exp\left(i eV\tau/\hbar\right) \sum_{\sigma_z} c_{b,\sigma_z}^+
  c_{D_y,\sigma_z} \\\nonumber&&+ h.c.\\
  {\cal H}_{T,D_x,c_1}&=&
  -\Sigma^{(2)}_{D_x,c_1} \sum_{\sigma_z} c_{{c_1},\sigma_z}^+
  c_{D_x,\sigma_z} + h.c.\\
  {\cal H}_{T,D_y,c_2}&=&
  -\Sigma^{(2)}_{D_y,c_2} \sum_{\sigma_z} c_{{c_2},\sigma_z}^+
  c_{D_y,\sigma_z} + h.c.
\label{eq:Tclosely-spaced-c2}
,
\end{eqnarray}
and we assume
\begin{equation}
  \label{eq:Sigma(2)}
  \Sigma^{(2)}_{D_x,a}=\Sigma^{(2)}_{D_y,b}
=\Sigma^{(2)}_{D_x,c_1} =\Sigma^{(2)}_{D_y,c_2}\equiv
\Sigma^{(2)}.
\end{equation}
We parameterize the contact transparency with
$\Gamma=\left(\Sigma^{(2)}\right)^2/W$.}

{In addition, we include the following ``crossed tunneling terms''
  between $D_x$ and $S_b,\,S_{c,2}$ and between $D_y$ and
  $S_a,\,S_{c,1}$:
\begin{eqnarray}
 \label{eq:Tclosely-spaced-b-crossed}
 {\cal H}_{T,D_x,b}(\tau)&=& -\Sigma^{(2)}_{D_x,b} \exp\left(i
 eV\tau/\hbar\right) \sum_{\sigma_z} c_{b,\sigma_z}^+
 c_{D_x,\sigma_z}\\&&\nonumber + h.c.\\ {\cal H}_{T,D_y,a}(\tau)&=&
 -\Sigma^{(2)}_{D_y,a} \exp\left(-i eV\tau/\hbar\right)
 \sum_{\sigma_z} c_{a,\sigma_z}^+ c_{D_y,\sigma_z}\\&&\nonumber +
 h.c.\\ {\cal H}_{T,D_x,c_2}&=& -\Sigma^{(2)}_{D_x,c_2}
 \sum_{\sigma_z} c_{{c_2},\sigma_z}^+ c_{D_x,\sigma_z} + h.c.\\ {\cal
   H}_{T,D_y,c_1}&=& -\Sigma^{(2)}_{D_y,c_1} \sum_{\sigma_z}
 c_{{c_1},\sigma_z}^+ c_{D_y,\sigma_z} + h.c.
\label{eq:Tclosely-spaced-c1-crossed},
\end{eqnarray}
and we assume
\begin{equation}
  \label{eq:Sigma-prime-2}
  \Sigma^{(2)}_{D_x,b}=\Sigma^{(2)}_{D_y,a}
=\Sigma^{(2)}_{D_x,c_2} =\Sigma^{(2)}_{D_y,c_1}\equiv
\Sigma'^{(2)}
.
\end{equation}
We parameterize the crossed contact transparency with
$\Gamma'=\left(\Sigma'^{(2)}\right)^2/W$.}

The Hamiltonian of the normal lead is given by Eq.~(\ref{eq:H-N}).
{Tunneling between the quantum dots $D_x,\,D_y$ and the normal lead
  $N$ is realized with the single-channel contacts $(D_x,N_x)$ and
  $(D_y,N_y)$:
  \begin{eqnarray}
  \label{eq:coupling-N1}
{\cal H}_{T,D_x,N_x}&=& -\Sigma^{(3)}_{D_x,N_x} \sum_{\sigma_z}
c_{N_x,\sigma_z}^+ c_{D_x,\sigma_z} + h.c.\\ {\cal H}_{T,D_y,N_y}&=&
-\Sigma^{(3)}_{D_y,N_y} \sum_{\sigma_z} c_{N_y,\sigma_z}^+
c_{D_y,\sigma_z} + h.c.  ,
\label{eq:coupling-N2}
\end{eqnarray}
  $N_x$ and $N_y$ being tight-binding sites belonging to the normal
  lead, and connected to $D_x$ and $D_y$ respectively by
  Eqs.~(\ref{eq:coupling-N1})-(\ref{eq:coupling-N2}). We additionally
  assume that $\Sigma^{(3)}_{D_x,N_x}$ and $\Sigma^{(3)}_{D_y,N_y}$
  are identical:
\begin{equation}
  \label{eq:Sigma-3}
  \Sigma^{(3)}_{D_x,N_x} = \Sigma^{(3)}_{N_x,D_x} =
  \Sigma^{(3)}_{D_y,N_y} = \Sigma^{(3)}_{N_y,D_y} \equiv \Sigma^{(3)}
  .
\end{equation}
}

This model will be used in the following sections~\ref{sec:2CPR},
\ref{sec:closing-Dyson}, \ref{sub2}, \ref{sec:specializing},
\ref{sec:transport} and \ref{sec:current-conservation}.

{\it Physical remarks:} Once disconnected from the superconducting or
normal leads, this phenomenological model of double 0D quantum dot
with normal lead in parallel naturally yields the pair of
opposite-energy levels that is relevant to the two-Cooper pair
resonance.

Now, we consider in section~\ref{sec:remote-0D} another
phenomenological {model that also includes} the
possibility of quartet current oscillating at the scale of the Fermi
wave-length as a function a dimension of the device.
\begin{figure}[htb]
  \includegraphics[width=.95\columnwidth]{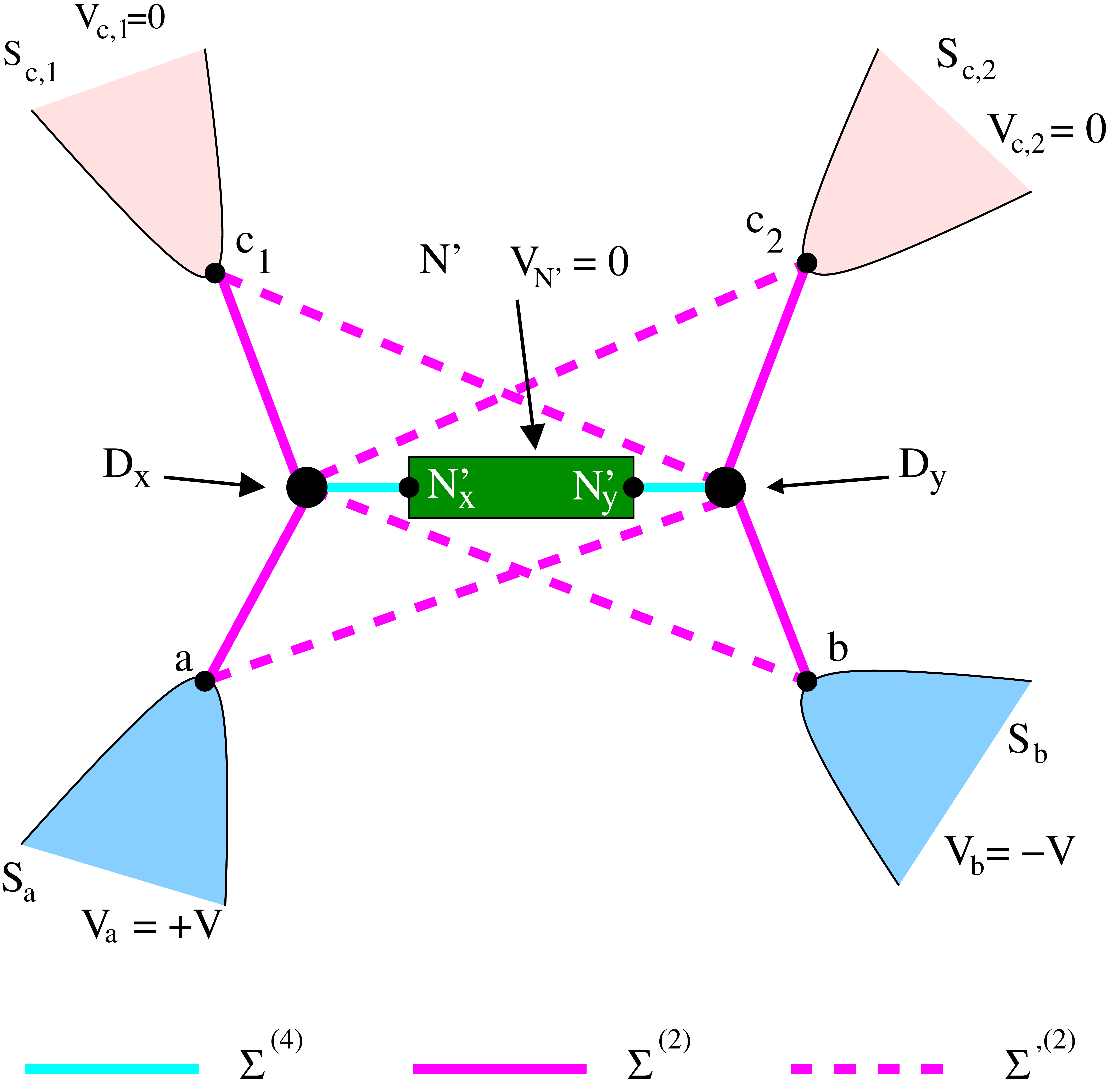}
  \caption{{\it The four-terminal phenomenological model of
      superconducting double 0D quantum dot with normal lead in
      series:} The figure shows a double 0D quantum dot connected to
    the four superconducting leads $S_a$, $S_b$, $S_{c,1}$ and
    $S_{c,2}$ biased at $V_{a,b}=\pm V$ and $V_{c,1}=V_{c,2}=0$, and
    to the four grounded normal leads $N_1$, $N_2$, $N_3$ and
    $N_4$. The corresponding Hamiltonian is provided in
    section~\ref{sec:remote-0D}. The coupling $\Sigma^{(4)}$ [see
      Eqs.~(\ref{eq:Sigma4-i})-(\ref{eq:Sigma4-f})] couples the two
    quantum dots $D_x$ and $D_y$ to the conductor $N'$ that, in the
    calculations, is characterized by the local and nonlocal Green's
    functions among the tight-binding sites $N_x$ and $N_y$. The
    ``direct'' coupling $\Sigma^{(2)}$ connects $D_x$ to $S_a$, $D_x$
    to $S_{c,1}$, $D_y$ to $S_b$ and $D_y$ to $S_{c,2}$, see
    Eqs.~(\ref{eq:Tclosely-spaced-a})-(\ref{eq:Sigma(2)}). The
    ``crossed'' coupling $\Sigma'^{(2)}$ connects $D_x$ to $S_b$,
    $D_x$ to $S_{c,2}$, $D_y$ to $S_a$ and $D_y$ to $S_{c,1}$, see
    Eqs.~(\ref{eq:Tclosely-spaced-b-crossed})-(\ref{eq:Sigma-prime-2}).
    This model will be used in the following
    sections~\ref{sec:closing-Dyson}, \ref{sub2},
    \ref{sec:specializing}, \ref{sec:transport}
    and~\ref{sec:num-2nodes}.
    \label{fig:geometrie-4T-double-dot-remote}. 
  }
\end{figure}

\subsubsection{Phenomenological model of double 0D
  quantum dot with normal lead in series}
\label{sec:remote-0D}

Now, we consider the Hamiltonian of a double 0D quantum dot with
normal lead in series, and connected to superconducting
leads. {We also provide a physical discussion. The
  numerical calculations presented in the forthcoming
  section~\ref{sec:num-2nodes} are based on this third stage of
  the simplifications.}

{\it The Hamiltonian:} {Namely, we consider a third model, i.e. a
  ``phenomenological model of double 0D quantum dot with normal lead
  in series'' connected by single-channel contacts through a 2D metal,
  see figure~\ref{fig:geometrie-4T-double-dot-remote}. The difference
  with the previous section~\ref{sec:closely-spaced-double} is in the
  coupling to the normal lead.}

As in the above section~\ref{sec:closely-spaced-double}, the quantum
dots $D_x$ and $D_y$ have the vanishingly small Hamiltonians of
Eqs.~(\ref{eq:H-dot-1}) and~(\ref {eq:H-dot-2}), {the superconducting
  leads $S_a$, $S_b$, $S_{c,1}$ and $S_{c,2}$ are described by the BCS
  Hamiltonian given by Eq.~(\ref{eq:H-BCS1}) and the normal lead is
  described by Eq.~(\ref{eq:H-N}).}  Tunneling between the leads
$S_a$, $S_b$, $S_{c,1}$ and $S_{c,2}$ and the double quantum dot
$(D_x,D_y)$ is described by the same single-channel contacts as in the
above
Eqs.~(\ref{eq:Tclosely-spaced-a})-(\ref{eq:Tclosely-spaced-c2}). In
addition, the crossed tunneling amplitudes given by
Eqs.~(\ref{eq:Tclosely-spaced-b-crossed})-(\ref{eq:Tclosely-spaced-c1-crossed})
are also included for completeness.

The coupling to the normal lead is phenomenologically accounted for by
replacing the tunneling amplitude $\Sigma^{(0)}$ between the quantum
dots by propagation through the normal conductor $N'$ connected
to the dots $D_x$ and $D_y$ by the single-channel tunneling amplitudes
$\Sigma^{(0)}_{D_x,N'_x}$ and $\Sigma^{(0)}_{D_y,N'_y}$ respectively:
\begin{eqnarray}
  \label{eq:Sigma4-i}
 {\cal H}_{T,D_x,N'_x}&=& -\Sigma^{(4)}_{D_x,N'_x} \sum_{\sigma_z}
  c_{N'_x,\sigma_z}^+ c_{D_x,\sigma_z} + h.c.\\
  {\cal H}_{T,D_y,N'_y}&=& -\Sigma^{(4)}_{D_y,N'_y} \sum_{\sigma_z}
  c_{N'_y,\sigma_z}^+ c_{D_y,\sigma_z} + h.c.
  ,
\end{eqnarray}
where we assume
\begin{equation}
  \label{eq:Sigma4-f}
  \Sigma^{(4)}_{D_x,N'_x}=\Sigma^{(4)}_{N'_x,D_x}=
  \Sigma^{(4)}_{D_y,N'_y}= \Sigma^{(4)}_{N'_y,D_y}\equiv \Sigma^{(4)}
  .
\end{equation}
The normal conductor $N'$ is characterized by the local Green's
functions at $N'_x$ or $N'_y$ , in addition to nonlocal ones from $N'_x$
to $N'_y$ or from $N'_y$ to $N'_x$ across $N'$. The nonlocal Green's
function between $N'_x$ and $N'_y$ generally has to be a complex number.

This model will be used in the following
sections~\ref{sec:closing-Dyson}, \ref{sub2}, \ref{sec:specializing},
\ref{sec:transport} and~\ref{sec:num-2nodes}.

{\it Physical remarks:} {Within this phenomenological model of double
  0D quantum dot with normal lead in series, the quartet critical
  current at resonance is sensitive to the Green's function connecting
  the two quantum dots $D_x$ and $D_y$. This Green's function is
  itself sensitive to the microscopic details of the model, i.e. on
  the value of the separation $R_0$ between the tight-binding sites
  $N'_x$ and $N'_y$ within the interval $\left[R'_0-\lambda_F/2,
    R'_0+\lambda_F/2\right]$, where $\lambda_F$ is the Fermi
  wave-length. Thus, this phenomenological model of double 0D quantum
  dot with normal lead in series produces strong sample-to-sample
  fluctuations of the quartet current, where different samples
  correspond to different values of $R_0$.

Once disconnected from the superconducting or normal leads, this
phenomenological model of double 0D quantum dot with normal lead in
series also yields the pair of energy levels that is central to the
two-Cooper pair resonance.

{In section~\ref{sec:num-2nodes}, we present numerical results for
  this model in the regime of strong sample-to-sample fluctuations. }
\begin{figure}[htb]
  \includegraphics[width=.9\columnwidth]{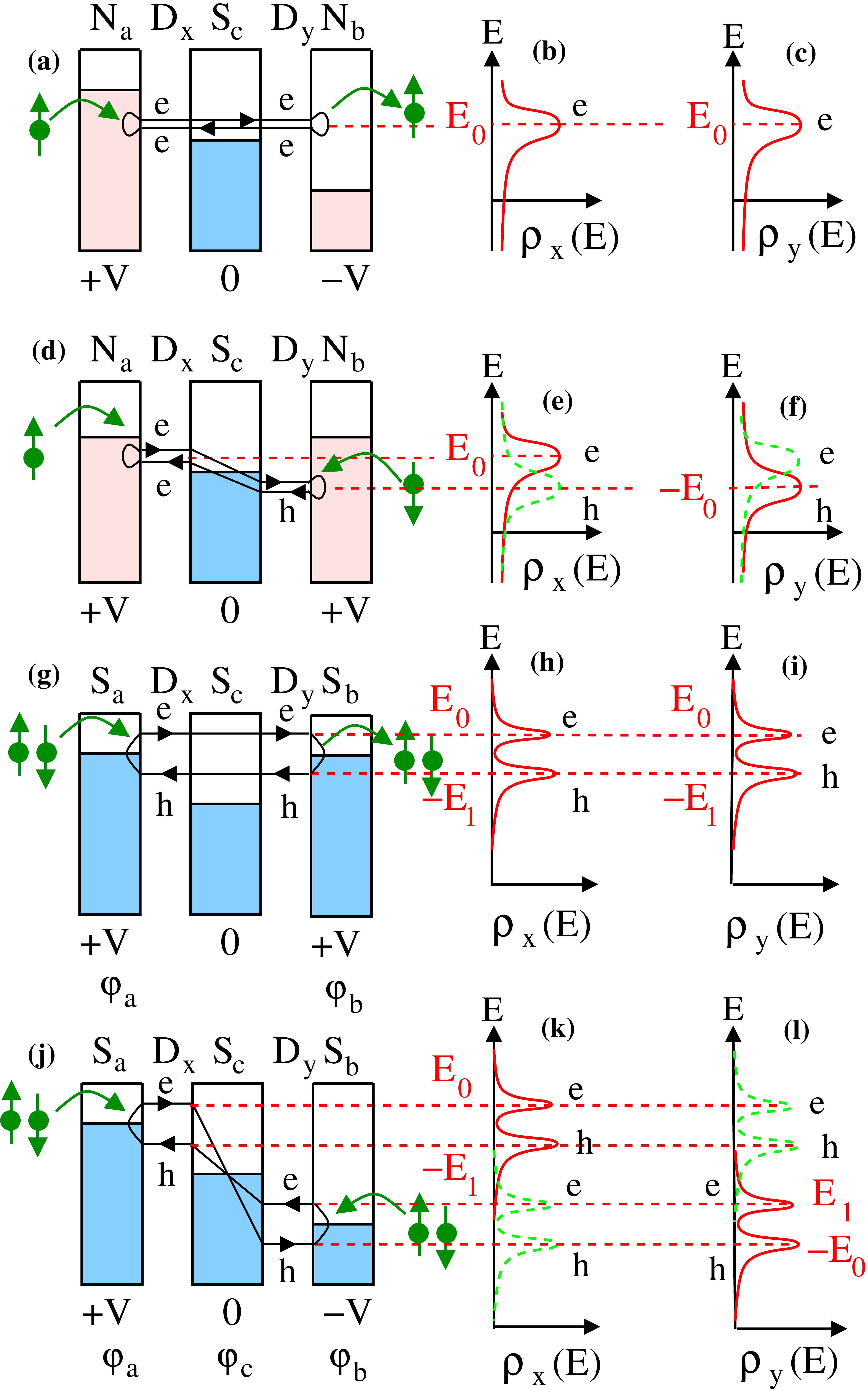}
  \caption{{\it The nonlocal resonances in three-terminal devices.}
    Panels a, d, g and j show the lowest-order diagrams of elastic
    cotunneling (EC), crossed Andreev reflection (CAR), double elastic
    cotunneling (dEC) and double crossed Andreev reflection (dCAR)
    respectively.  Panels b, e, h and k show the energy-$E$-dependence
    of the density of states $\rho_x(E)$ on the quantum dot $D_x$ and
    panels c, f, i and l show the corresponding $\rho_y(E)$ on the
    quantum dot $D_y$. Panels e, f, k and l also show for completeness
    by green dashed lines the complementary resonances that do not
    directly contribute to the corresponding transport processes.
  \label{fig:presentation2}
  }
\end{figure}

\subsubsection{{Further physical remarks on the choice of a model}}
\label{sec:comments-choice}
Now, we explain why we focus on the double quantum dot
Hamiltonians presented in the above
subsections~\ref{sec:closely-spaced-double} and~\ref{sec:remote-0D},
instead of a single, three or four-quantum dot Hamiltonians.

First, we comparatively examine single and double 0D quantum dots.

At equilibrium, the ABS of a single 0D quantum dot have energy scale
$\Gamma$ typically set by $\Gamma=\Sigma^2/W$, where $\Sigma$ is the
tunneling amplitude between the dot and the superconducting leads. It
turns out that, at small $\Gamma$, the quasiadiabatic regime $V\alt
V_*$ with $eV_*\sim \Gamma$, does not produce enhanced quartet
critical current according to the forthcoming
subsection~\ref{sec:transport} where the two-Cooper pair resonance is
put in correspondence with nonadiabatic effects.

At equilibrium, double 0D quantum dots are characterized by levels at the
opposite energies $\pm \Omega$, where $\Omega$ is set by the tunneling
amplitude $\Sigma^{(0)}$ between the dots, see
Eq.~(\ref{eq:H-xy}}). Then, the voltage energy $eV_*=\Omega$ of the
first resonance remains finite in the weak-coupling limit, which
implies nonadiabatic behavior and enhancement of the quartet critical
current at low $V$ if $\Omega$ is small compared to the
superconducting gap $\Delta$.

Double 0D quantum dots are thus better candidates than single 0D
quantum dots to produce large quartet critical current at small bias
voltage because the two-Cooper pair resonance is also there in the
weak-coupling regime of {small-$\Gamma$, see also the
  forthcoming section~\ref{sec:low-V}.}

The interest of double 0D quantum dots is also related to the
observation that the corresponding four energy levels can accommodate
the four fermions of a quartet as a ``real state''. In this sense, the
double 0D quantum dots have the ``minimal complexity'' in comparison
with three- or four-quantum dot devices.

\section{Mechanism for the two-Cooper pair resonance}
\label{sec:mechanism}

In this section, we present a general mechanism for emergence of the
two-Cooper pair resonance.  We start in subsection~\ref{sec:2CPR} with
introducing the three-terminal $N_a$-dot-$S_c$-dot-$N_b$ Cooper pair
beam splitters, and the $S_a$-dot-$S_c$-dot-$S_b$ three-terminal
Josephson junctions. Next, we calculate effective non-Hermitian
self-energies and Hamiltonians in subsection~\ref{sec:non-her}. The
transport formula are discussed in subsection~\ref{sec:transport} in
connection with emergence of resonances. {Current
  conservation is discussed in
  subsection~\ref{sec:current-conservation}.}

\subsection{Three-terminal two-Cooper pair resonance}
\label{sec:2CPR}

In this subsection, we introduce the two-Cooper pair resonance in
three-terminal Josephson junctions, i.e. in $S_a$-dot-$S_c$-dot-$S_b$
devices, starting with nonlocal resonances in the three-terminal
$N_a$-dot-$S_c$-dot-$N_b$.

Figures~\ref{fig:presentation2}a, \ref{fig:presentation2}b and
\ref{fig:presentation2}c show the condition for nonlocal resonance of
elastic cotunneling (EC)
\cite{Yeyati2007,theory-CPBS1,theory-CPBS2,theory-CPBS3,theory-CPBS4,theory-CPBS4-bis,theory-CPBS4-ter,theory-CPBS5,theory-CPBS6,theory-CPBS7,theory-CPBS8,theory-CPBS9,theory-CPBS10,theory-CPBS11}
in $N_a$-dot-$S_c$-dot-$N_b$ Cooper pair beam splitters. EC transfers
single-particle states from $N_a$ to $N_b$ across $S_c$, and
contributes to negative nonlocal conductance ${\cal G}_{a,b}=\partial
I_a/\partial V_b$ on the voltage biasing condition $V_a=-V_b$, where
$I_a$ is the current transmitted into the normal lead $N_a$.  EC is
resonant if both quantum dots $D_x$ and $D_y$ have levels at the same
energy $\epsilon_x=\epsilon_y$, where $\omega=\epsilon_x=\epsilon_y$
is the energy of the incoming electron in
figure~\ref{fig:presentation2}a.

Figures~\ref{fig:presentation2}d, \ref{fig:presentation2}e,
\ref{fig:presentation2}f show CAR
\cite{Yeyati2007,theory-CPBS1,theory-CPBS2,theory-CPBS3,theory-CPBS4,theory-CPBS4-bis,theory-CPBS4-ter,theory-CPBS5,theory-CPBS6,theory-CPBS7,theory-CPBS8,theory-CPBS9,theory-CPBS10,theory-CPBS11}
in $N_a$-dot-$S_c$-dot-$N_b$ Cooper pair beam splitters, which
reflects by nonlocal Andreev reflection spin-up electron impinging
from $N_a$ as spin-down hole transmitted into $N_b$, leaving a Cooper
pair in the central $S_c$. CAR contributes for positive value to the
nonlocal conductance ${\cal G}_{a,b}$ at the bias voltages $V_a=V_b$
and $V_c=0$. The nonlocal CAR resonance is obtained if the quantum
dots $D_x$ and $D_y$ have levels at the opposite energies $\epsilon_x
= -\epsilon_y \equiv \omega$.

Considering $S_a$-dot-$S_c$-dot-$S_b$ three-terminal Josephson
junctions, figures \ref{fig:presentation2}g, \ref{fig:presentation2}h,
\ref{fig:presentation2}i feature double elastic cotunneling (dEC)
\cite{theory-CPBS8,theory-CPBS9,theory-CPBS10}, which transfers Cooper
pairs from $S_a$ to $S_b$ across $S_c$ at the bias voltages $V_a=V_b$.
This process is resonant if $\omega=E_0$ (for the spin-up electron
crossing $D_x$ or $D_y$) and $-\omega+2eV=-E_1$ (for the spin-down
hole crossing $D_x$ or $D_y$). Thus, dEC is resonant if
$eV=(E_0-E_1)/2$ and $\omega=E_0$.

Concerning double crossed Andreev reflection (dCAR) in
$S_a$-dot-$S_c$-dot-$S_b$ three-terminal Josephson junctions in
figures \ref{fig:presentation2}j, \ref{fig:presentation2}k and
\ref{fig:presentation2}l, two Cooper pairs from $S_a$ and $S_b$ biased
at the voltages $\pm V$ cooperatively enter the grounded $S_c$,
producing transient correlations among four fermions, i.e. the
so-called quartets. Then,
\begin{eqnarray}
  \label{eq:A1}
  \omega&=&E_0\\
  -\omega+2eV&=&-E_1
  \label{eq:B1}
\end{eqnarray}
are obtained at resonance. Conversely, the same
Eqs.~(\ref{eq:A1})-(\ref{eq:B1}) are obtained for resonance of the
spin-up electron and spin-down holes crossing $D_y$.

Now, we assume that $D_x$ and $D_y$ are gathered into a single
multilevel quantum dot. Assuming the opposite energies $E_0=-E_1\equiv
\Omega$ implies $\omega=\Omega$ and $-\omega+2eV=\Omega$, which yields
\begin{equation}
  \label{eq:resonance-condition}
  eV=\omega=\Omega
\end{equation}
for the nonlocal quartet resonance.

Overall, the argument leading to Eq.~(\ref{eq:resonance-condition})
confirms that nonlocal quartet resonance is produced at voltage energy
$eV$ that can be much smaller than the superconducting gap $\Delta$,
if the energy scales $\pm \Omega$ are also within {$\pm\Delta$, see
  also the preceding subsection~\ref{sec:comments-choice}.}

\subsection{Effective non-Hermitian self-energy and Hamiltonian}
\label{sec:non-her}

In this subsection, we present how effective non-Hermitian
self-energies or Hamiltonians can be obtained in the infinite-gap
limit from the models presented in the above
section~\ref{sec:H}.

We start in section~\ref{sec:non-her} with the Dyson equations for the
three models presented in the above
subsections~\ref{sec:multilevel-quantum-dot},
\ref{sec:closely-spaced-double} and \ref{sec:remote-0D}.  Emergence of
non-Hermitian self-energy in the infinite-gap limit is discussed in
subsection~\ref{sub2}. Non-Hermitian Hamiltonians are next presented
for double quantum dots in the infinite-gap limit, see
subsection~\ref{sec:specializing}.

\subsubsection{Closing the Dyson equations}
\label{sec:closing-Dyson}
In this subsection, we provide the starting-point Dyson equations. We
adopt a general viewpoint on the three models of
subsections~\ref{sec:multilevel-quantum-dot},
\ref{sec:closely-spaced-double} and~\ref{sec:remote-0D}, i.e.
multilevel quantum dots, double 0D quantum dots connected in parallel
and in series respectively, see also
figures~\ref{fig:geometrie-4T-multilevel},
\ref{fig:geometrie-4T-double-dot}
and~\ref{fig:geometrie-4T-double-dot-remote}.

The ``quantum dot $D$'' is coupled to $p_0$ superconducting leads
labeled by $n_p=1,\,...,\,p_0$ and to $q_0$ normal leads labeled by
$n_q=1\,,...,\,q_0$, where $p_0=q_0=4$ are used in
figure~\ref{fig:geometrie-4T-multilevel}. The Dyson equations take the
form
\begin{eqnarray}
  \nonumber \hat{G}_{D,D}&=& \hat{g}_{D,D}
  +\sum_{n_q=1}^{q_0}\sum_{n_{q'}=1}^{q_0} \hat{g}_{D,D}
  \hat{\Sigma}^{(5)}_{D,N_{n_q}} \hat{g}_{N_{n_q},N_{n_{q'}}}
  \hat{\Sigma}^{(5)}_{N_{n_{q'}},D} \hat{G}_{D,D}
  \\&&+\sum_{n_p=1}^{p_0} \hat{g}_{D,D} \hat{\Sigma}^{(5)}_{D,S_{n_p}}
  \hat{g}_{S_{n_p},S_{n_p}} \hat{\Sigma}^{(5)}_{S_{n_p},D}
  \hat{G}_{D,D} .\label{eq:GDD-1}
\end{eqnarray}
The notation ``$D$'' in Eq.~(\ref{eq:GDD-1}) refers to the collection
of the tight-binding sites in absence of coupling to the
superconducting or normal leads, and to the direct coupling between
them. For instance, $D$ represents finite-size tight-binding
multilevel quantum dot, see
subsection~\ref{sec:multilevel-quantum-dot}, or $D_x$-$D_y$ double
quantum dot, see subsections~\ref{sec:closely-spaced-double}
and~\ref{sec:remote-0D}. The notation $\Sigma^{(5)}$ denotes generic
hopping amplitude, with the normal lead $N_{n_q}$ (corresponding to
$\hat{\Sigma}^{(5)}_{D,N_{n_q}} = \hat{\Sigma}^{(5)}_{N_{n_q},D}$), or
with the superconducting lead $S_{n_p}$ (corresponding to
$\hat{\Sigma}^{(5)}_{D,S_{n_p}} = \hat{\Sigma}^{(5)}_{S_{n_p},D}$).

The notations $\hat{g}_{N_{n_q},N_{n_{q'}}}$ and
$\hat{g}_{S_{n_p},S_{n_p}}$ stand for the bare Green's functions of
the normal or superconducting leads $N_{n_q}$, $N_{n_{q'}}$ or
$S_{n_p}$ respectively, and $\hat{g}_{D,D}$, $\hat{G}_{D,D}$ are the
bare and fully dressed quantum dot Green's functions. The bare Green's
functions are given by
\begin{eqnarray}
  \label{eq:AA1}
  \hat{g}_{D,D,{\bf x}, {\bf y}}^{A,1,1}
  &=& \sum_\psi \langle {\bf x}|\psi\rangle
  \frac{1}{\omega-\epsilon_\psi-i\eta}
  \langle \psi|{\bf y}\rangle\\
  \hat{g}_{D,D,{\bf x}, {\bf y}}^{A,2,2}
  &=& \sum_\psi \langle {\bf x}|\psi\rangle
  \frac{1}{\omega+\epsilon_\psi-i\eta}
  \langle \psi|{\bf y}\rangle
  ,
  \label{eq:AA2}
\end{eqnarray}
{where $\omega$ is the real-part of the energy and
  ``$1,1$'',``$2,2$'' refer to the electron-electron and hole-hole
  channels respectively.} In addition, $\hat{g}_{D,D,{\bf x}, {\bf
    y}}^{A,1,2}=\hat{g}_{D,D,{\bf x}, {\bf y}}^{A,2,1}=0$ because of
the absence of superconducting pairing on the quantum dot. The Dynes
parameter $\eta$ in Eqs.~(\ref{eq:AA1})-(\ref{eq:AA2})
\cite{Kaplan,Dynes,Pekola1,Pekola2} makes the distinction between the
advanced and retarded Green's functions, and it can be used to capture
relaxation on the quantum dot~\cite{FWS,Bergeret}. As in the previous
Eq.~(\ref{eq:H-multi}), the notation $|\psi\rangle$ in
Eqs.~(\ref{eq:AA1})-(\ref{eq:AA2}) stands for the single-particle
states, and $\langle {\bf x}|\psi\rangle$, $\langle {\bf
  y}|\psi\rangle$ are the corresponding wave-functions at the
tight-binding sites ${\bf x}$ and~${\bf y}$.

The bare and fully dressed Green's functions
$\hat{g}_{D,D}$ and $\hat{G}_{D,D}$ have entries in the Nambu labels,
and in the tight-binding sites making the contacts between the dot and
the normal or superconducting leads $N_{n_q}$ or $S_{n_p}$. In
addition, those matrices have entries in the set of the harmonics of
the Josephson frequency.

\subsubsection{Effective non-Hermitian self-energy}
\label{sub2}
In this subsection, we explain how effective non-Hermitian self-energy
is obtained in the infinite-gap limit. {It was already
  mentioned in the introductory section that the infinite-gap limit}
was introduced and considered over the last few years, see for
instance Refs.~\onlinecite{topo1_plus_Floquet,Zazunov,Meng} to cite
but a few.
  
Specifically, Eq.~(\ref{eq:GDD-1}) is rewritten as
\begin{eqnarray}
    \label{eq:Dy2}
&&
    \left[\hat{g}_{D,D}^{-1}(\omega)-\sum_{n_q=1}^{q_0}\sum_{n_{q'}=1}^{q_0}
      \hat{\Sigma}^{(5)}_{D,N_{n_q}} \hat{g}_{N_{n_q},N_{n_{q'}}}
      \hat{\Sigma}^{(5)}_{N_{n_{q'}},D} \right.\\&&-\left.
      \sum_{n_p=1}^{p_0} \hat{\Sigma}^{(5)}_{D,S_{n_p}}
      \hat{g}_{S_{n_p},S_{n_p}} \hat{\Sigma}^{(5)}_{S_{n_p},D} \right]
    \hat{G}_{D,D}(\omega)= \hat{I} \nonumber .
\end{eqnarray}
In addition, $\hat{g}_{S,S}$ is independent on $\omega$ in the
considered infinite-gap limit, and the local Green's function
$\hat{g}_{N_{n_q},N_{n_{q'}}}$ in the normal leads is also taken as
being independent on energy.

Then, Eq.~(\ref{eq:Dy2}) can be rewritten as
\begin{eqnarray}
  \left[\omega-\hat{\Sigma}_{Non\,Her.}(\omega) \right]^{-1} \hat{G}_{D,D}(\omega)= \hat{I}
  ,
\end{eqnarray}
where the self-energy
\begin{eqnarray}
  \label{eq:self-energy-non-her}
  \hat{\Sigma}_{Non\,Her.}(\omega)&=&\omega-\hat{g}_{D,D}^{-1}(\omega)\\\nonumber
  &&+\sum_{n_q=1}^{q_0}\sum_{n_{q'}=1}^{q_0}
  \hat{\Sigma}^{(5)}_{D,N_{n_q}} \hat{g}_{N_{n_q},N_{n_{q'}}}
  \hat{\Sigma}^{(5)}_{N_{n_{q'}},D} \\&&+ \sum_{n_p=1}^{p_0}
  \hat{\Sigma}^{(5)}_{D,S_{n_p}} \hat{g}_{S_{n_p},S_{n_p}}
  \hat{\Sigma}^{(5)}_{S_{n_p},D} \nonumber
\end{eqnarray}
in non-Hermitian, due to $\hat{g}_{N_{n_q},N_{n_{q'}}}$.

\subsubsection{Specializing to double quantum dots}
\label{sec:specializing}
In this subsection, we discuss emergence of non-Hermitian Hamiltonian
for the phenomenological model of double 0D quantum dots. The normal
leads in parallel, see subsection~\ref{sec:closely-spaced-double}, are
treated in the same framework as normal leads in series, see
subsection~\ref{sec:remote-0D}.

{The retarded and advanced Green's functions $\sim
  1/(\omega-{\cal H}_{eff}^R)$ and $\sim 1/(\omega-{\cal H}_{eff}^A)$
  are obtained from the effective non-Hermitian Hamiltonians ${\cal
    H}_{eff}^R\equiv {\cal H}_{eff}$ and ${\cal H}_{eff}^A=\left({\cal
    H}_{eff}\right)^+$ respectively. Their complex eigenvalues receive
  interpretation of the real and imaginary parts of the ABS energies,
  yielding coherent oscillations and damping in the ABS dynamics,
  respectively.}

We make use of the same compact notations as in the previous
subsections~\ref{sec:closing-Dyson} and~\ref{sub2}. Then we show that
the effective non-Hermitian self-energy
$\hat{\Sigma}_{Non\,Her.}(\omega)$ in
Eq.~(\ref{eq:self-energy-non-her}) becomes energy-independent for
those double 0D quantum dots, thus defining the effective
non-Hermitian Hamiltonian $\hat{\cal H}_{eff}^{(\infty)}\equiv
\hat{\Sigma}_{Non\,Her.}$ in the infinite-gap limit.

{The matrices in Eq.~(\ref{eq:Dy2}) have entries in the tight-binding
  sites at the boundary of the quantum dot, where the tunneling
  self-energy between the dot and the superconducting or normal leads
  is acting. The boundary is identical to the bulk for the double 0D
  quantum dots $D_x$-$D_y$ isolated from the leads. Thus, $g_{D,D}$
  takes the value
  \begin{equation}
    \label{eq:gD-D}
  \hat{g}_{D,D}^R=\left(\omega+i\eta-\hat{\cal H}_{dot}\right)^{-1}
  ,
  \end{equation}
  where $\hat{\cal H}_{dot}$ is the double 0D quantum dot
  Hamiltonian. The retarded Green's functions of the double 0D quantum
  dots with normal leads in parallel, see
  figure~\ref{fig:geometrie-4T-double-dot}, can then be expressed as
\begin{equation}
  \hat{G}_{D,D}^R= \left(\omega+i\eta-\hat{\cal H}_{eff}^{(\infty)}\right)^{- 1}
  ,
\end{equation}
where the infinite-gap-limit effective Hamiltonian
\begin{eqnarray}
  \label{eq:H-infinite}
  \hat{\cal H}_{eff}^{(\infty)}&=& \hat{\cal H}_{dot} +\sum_{n_q=1}^{q_0}\sum_{n_{q'}=1}^{q_0}
  \hat{\Sigma}^{(5)}_{D,N_{n_q}} \hat{g}_{N_{n_q},N_{n_{q'}}}
  \hat{\Sigma}^{(5)}_{N_{n_{q'}},D} \\&&+\sum_{n_p=1}^{p_0}
  \hat{\Sigma}^{(5)}_{D,S_{n_p}} \hat{g}_{S_{n_p},S_{n_p}}
  \hat{\Sigma}^{(5)}_{D_{n_p},D} \nonumber
\end{eqnarray}
is non-Hermitian, due to $\hat{g}_{N_{n_q},N_{n_{q'}}}$.}

{We note that infinite-gap Hamiltonians can more
  generally be deduced from the self-energy
  $\hat{\Sigma}_{Non\,Her.}(\omega)$ in
  Eq.~(\ref{eq:self-energy-non-her}) by extending the domain of
  definition of the tunneling amplitude to all tight-binding sites of
  the quantum dots, and associating vanishingly small tunneling
  amplitudes to those fictitious superconducting leads.}

\subsection{Emergence of resonances in the transport formula}
\label{sec:transport}

{In this subsection, we demonstrate that the current at resonance is
  inverse-proportional to the damping rate.}

Using the same notations as in the previous
subsections~\ref{sec:closing-Dyson}, \ref{sub2}
and~\ref{sec:specializing}, the current flowing from the multilevel
quantum dot into the superconducting lead $S_{n_p}$ is given by the
following energy-integral \cite{AVERIN,Cuevas,Caroli,Yeyati-noise-equilibrium}:
\begin{eqnarray}
  \label{eq:I-N-nu}
  I_{D,S_{n_p}}&=&\frac{e}{\hbar} \int d\omega \times\\&& \mbox{Tr}
  \left\{\hat{\sigma}_N^z \left[ \hat{\Sigma}_{D,S_{n_p}}^{(5)}
    \hat{G}^{+,-}_{S_{n_p},D} - \hat{\Sigma}_{S_{n_p},D}^{(5)}
    \hat{G}^{+,-}_{D,S_{n_p}}\right]\right\} ,
  \nonumber
  \end{eqnarray}
where the Pauli matrix $\hat{\sigma}_N^z$ is defined as
$\hat{\sigma}_N^z=\mbox{diag}(1,-1)$ on the set of the Nambu labels,
which is referred to as the subscript ``$N$'' for
``Nambu''. {In Eq.~(\ref{eq:I-N-nu}), $\hat{G}^{+,-}$
  is the fully dressed Keldysh Green's function, see
  Refs.~\onlinecite{Caroli,Cuevas,Yeyati-noise-equilibrium}.} The bare
superconducting Green's functions $\hat{g}^{+,-}_{S_{n_p},S_{n_p}}=0$
are vanishingly small if the infinite-gap limit is taken, and
$\hat{g}^{+,-}_{N_{n_q},N_{n_q'}}\ne 0$ is nonvanishingly small in the
normal leads, due the corresponding finite value of the normal density
of states.

\begin{figure*}[htb]
    \includegraphics[width=.8\textwidth]{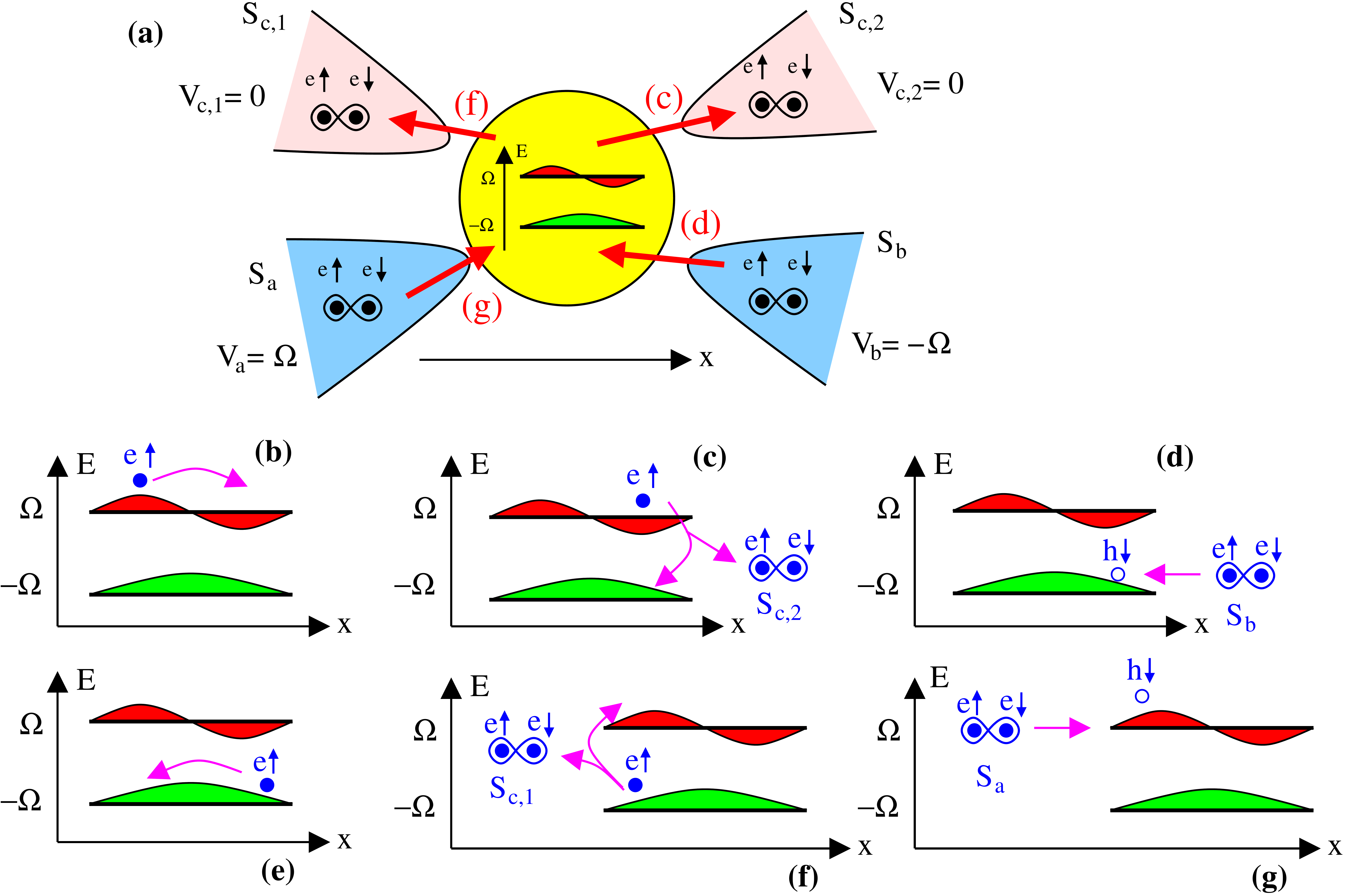} 
    \caption{Panel a schematically shows a quantum dot supporting
      levels at the opposite energies $E=\pm \Omega$, coupled to the
      superconducting leads $S_a$ and $S_b$ biased at $eV_{a,b}=\pm
      \Omega$ and to the grounded $S_{c,1}$ and $S_{c,2}$. Panels b-g
      represent two levels at the opposite energies $\pm \Omega$,
      together with the corresponding wave-functions, with absence of
      node at $E=-\Omega$, and a node at $E=\Omega$. Panels b-g show
      the time-evolution of the quartet process, involving Cooper pair
      transmitted into $S_{c,2}$ (panel c), Cooper pair taken from
      $S_b$ (panel d), Cooper transmitted into $S_{c,1}$ (panel f) and
      Cooper pair taken from $S_a$ (panel g). Detailed comments about
      panels b-g are given in the text.
  \label{fig:evolution}
  }
\end{figure*}

The two terms $\hat{\Sigma}_{D,S_{n_p}}^{(5)}
\hat{G}^{+,-}_{S_{n_p},D}$ and $\hat{\Sigma}_{S_{n_p},D}^{(5)}
\hat{G}^{+,-}_{D,S_{n_p}}$ in Eq.~(\ref{eq:I-N-nu}) can be expanded as follows:
\begin{eqnarray}
  \label{eq:DK1}&&
\hat{\Sigma}_{D,S_{n_p}}^{(5)} \hat{G}^{+,-}_{S_{n_p},D}=\\\nonumber&&
\sum_{q,q'} \hat{\Sigma}_{D,S_{n_p}}^{(5)} \left[\hat{I}+\hat{G}^R
  \hat{\Sigma}^{(5)}\right]_{S_{n_p},N_q} \hat{g}^{+,-}_{N_q,N_{q'}}
\left[\hat{I}+\hat{\Sigma}^{(5)}
  \hat{G}^A\right]_{N_{q'},D}\\ \label{eq:DK2}&&\hat{\Sigma}_{S_{n_p},D}^{(5)}
\hat{G}^{+,-}_{D,S_{n_p}}=\\&& \nonumber\sum_{q,q'}
\hat{\Sigma}_{S_{n_p},D}^{(5)} \left[\hat{I}+\hat{G}^R
  \hat{\Sigma}^{(5)}\right]_{D,N_q} \hat{g}^{+,-}_{N_q,N_{q'}}
\left[\hat{I}+\hat{\Sigma}^{(5)} \hat{G}^A\right]_{N_{q'},S_{n_p}} ,
\end{eqnarray}
where $N_q$ and $N_{q'}$ run over the interfaces between the quantum
dot and the normal lead.

Then, Eqs.~(\ref{eq:I-N-nu})-(\ref{eq:DK2}) become
\begin{eqnarray}
  \label{eq:I-N-nu3}
  I_{D,S_{n_p}}&=&\frac{e}{\hbar} \int d\omega \times\\&& \mbox{Tr}
  \left\{\hat{\sigma}_N^z \left[ \hat{\Sigma}_{D,S_{n_p}}^{(5)}
    \hat{g}^R_{S_{n_p},S_{n_p}} \hat{\Sigma}_{S_{n_p},D}^{(5)}
    \hat{G}^R_{D,D}\hat{\Gamma}_{D,D}^{+,-,N}
    \hat{G}^A_{D,D}\right.\right.\nonumber\\
    &&- \left.\left. \hat{\Sigma}_{S_{n_p},D}^{(5)}
    \hat{G}^R_{D,D} \hat{\Gamma}_{D,D}^{+,-,N} \hat{G}^A_{D,D} \hat{\Sigma}_{D,S_{n_p}}^{(5)}
    \hat{g}^R_{S_{n_p},S_{n_p}}\right]\right\} ,
  \nonumber
\end{eqnarray}
where
\begin{equation}
  \hat{\Gamma}_{D,D}^{+,-,N}= \sum_{n_q=1}^{q_0} \sum_{n_{q'}=1}^{q_0}
  \hat{\Sigma}^{(5)}_{D,N_{n_q}} \hat{g}^{+,-}_{N_{n_q},N_{n_{q'}}}
  \hat{\Sigma}^{(5)}_{N_{n_{q'}},D}
\end{equation}
is diagonal in Nambu and in Floquet.

The retarded and advanced Green's functions are approximated as
\cite{engineering,Berry}
\begin{eqnarray}
  \label{eq:resolvent1}
  \hat{G}^R_{D,D}&\simeq&
  \sum_{k',l'}
  \frac{\hat{R}_{k',l'}^R}{\omega-k'eV-E_{l'}+i\delta_{l'}}\\
  \hat{G}^A_{D,D}&\simeq&
  \sum_{k'',l''}
  \frac{\hat{R}_{k'',l''}^A}{\omega-k''eV-E_{l''}-i\delta_{l''}}
  \label{eq:resolvent2}
  ,
\end{eqnarray}
where $k'$, $l'$, $k''$ and $l''$ are four integers, $E_{l'}$,
$E_{l''}$ and $\delta_{l'}$, $\delta_{l''}$ are the Floquet energies
and line-width broadening, and $\hat{R}_{k',l'}^R$,
$\hat{R}_{k'',l''}^A$ are the matrix residues. Then, inserting
Eqs.~(\ref{eq:resolvent1})-(\ref{eq:resolvent2}) into
Eq.~(\ref{eq:I-N-nu3}) and integrating over the energy $\omega$ yields
\begin{eqnarray}
  \label{eq:delta_l}
  I_{D,S_{n_p}}&=&\frac{\pi e}{2\hbar} 
  \sum_{(k,l)}\frac{1}{\delta_l}\mbox{Tr}
  \left\{\hat{\sigma}_N^z \times\right.\\
    &&\left.\left[
      \hat{\Sigma}_{D,S_{n_p}}^{(5)}
      \hat{g}_{S_{n_p},S_{n_p}} \hat{\Sigma}_{S_{n_p},D}^{(5)},
      \hat{R}_{k,l}^R \hat{\Gamma}_{D,D}^{+,-,N} \hat{R}_{k,l}^A\right]_-\right\}
  ,
  \nonumber
\end{eqnarray}
where $\left[...,...\right]_-$ is a commutator and $\sum_{(k,l)}$ denotes
summation over the pairs of labels $k$ and $l$ such that
\begin{equation}
  k'eV+E_{l'}=k''eV+E_{l''}\equiv k eV+E_l
  .
\end{equation}
Thus, we find that
the current is inverse proportional to the damping rate set by the
parameter $\delta_l$.

\subsection{Current conservation}
\label{sec:current-conservation}

Now, we consider that the effective non-Hermitian Hamiltonian with
complex eigenvalues originates from the Hermitian Hamiltonians
presented in the above section~\ref{sec:H}, see also
Appendix~\ref{sec:evolution-t-0-app}. It turns out that the total
current is conserved once the fraction of the current transmitted into
the normal lead has been taken into account. Current conservation can
be demonstrated by assuming that the average number of fermions
$\langle\hat{N}_{dot}\rangle$ on the quantum dot is stationary. The
Hamiltonian ${\cal H}$ can be written as a sum of the Hamiltonians of
the quantum dot, of the lead, and tunneling between them. Then, $d
\langle\hat{N}_{dot}\rangle/dt = 0$ is equivalent to current
conservation. Thus, non-Hermitian effective Hamiltonian does not
contradict current conservation. {We note that
  self-consistent algorithms were used in Ref.~\onlinecite{FWS} to
  impose current conservation in three-terminal Josephson junctions in
  the presence of a phenomenological Dynes parameter $\eta$.}

\section{Mechanism for the inversion}
\label{sec:mecha-inversion}

We discussed in the previous section~\ref{sec:mechanism} how sharp
resonance peaks can appear in the voltage-dependence of the quartet
critical current. Now, we provide simple arguments for the magnetic
flux-$\Phi$ sensitivity of the quartet critical current in the $V=0^+$
adiabatic limit, focusing on how the quartet critical current
$I_{q,c}(\Phi=0)$ in zero field $\Phi=0$ compares to
$I_{q,c}(\Phi=\pi)$ at half-flux quantum $\Phi=\pi$. Inversion
corresponds to larger quartet critical current at $\Phi=\pi$ than at
$\Phi=0$, i.e. $I_{q,c}(\Phi=\pi)>I_{q,c}(\Phi=0)$.  Absence of
inversion corresponds to $I_{q,c}(\Phi=\pi)<I_{q,c}(\Phi=0)$.

Figure~\ref{fig:evolution} shows a sequence of microscopic processes
for the quartets in the presence of the two energy levels $E=\pm
\Omega$ on the quantum dot. Figure~\ref{fig:evolution}b shows spin-up
electron at energy $E=\Omega$ on the left-part of the two-level
quantum dot, and how it moves to the
right-part. Figure~\ref{fig:evolution}c shows spin-up electron at
energy $E=\Omega$ on the right part of the junction, and how it is
converted by Andreev reflection into spin-down hole at energy
$E=-\Omega$ at the interface with the $S_{c,2}$ superconducting
lead. Figure~\ref{fig:evolution}d shows the resulting spin-down hole
at energy $E=-\Omega$, together with a Cooper pair taken from
$S_b$. Figure~\ref{fig:evolution}e shows the resulting spin-up
electron at energy $E=-\Omega$ on the right-part of the quantum dot,
and how it moves to the left part. Figure~\ref{fig:evolution}f shows
spin-up electron at energy $E=-\Omega$ on the left part of the
junction, and how it is converted into spin-down hole at energy
$E=\Omega$ by Andreev reflection at the interface with
$S_{c,1}$. Figure~\ref{fig:evolution}g shows the resulting spin-down
hole on the left-part of the junction at the energy $E=\Omega$,
together with absorption of a Cooper pair taken from $S_a$, thus
coming back to the initial state in figure~\ref{fig:evolution}b.

\begin{figure*}[htb]
  \includegraphics[width=.9\textwidth]{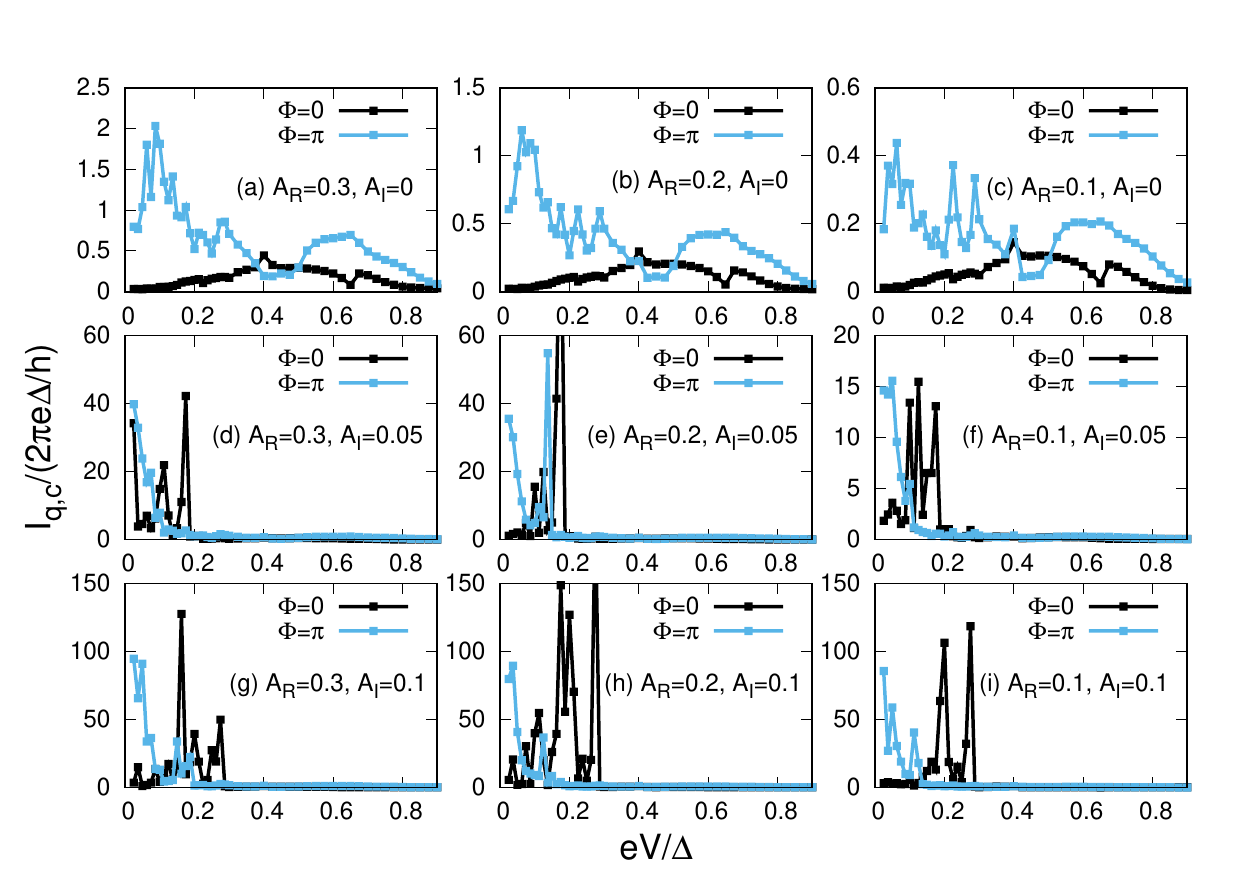}
  \caption{{\it The $eV/\Delta$-dependence of the quartet critical
      current.} Each panel of this figure shows
    $I_{q,c}(eV/\Delta,\Phi=0)$ and $I_{q,c}(\Phi=\pi,eV/\Delta)$ as a
    function of the normalized bias voltage $eV/\Delta$, for $A_I=0$
    and $A_R=0.3/W$ (a), $A_R=0.2/W$ (b) and $A_R=0.1/W$ (c). Panels
    d, e and f correspond to $A_I=0.05/W$ and $A_R=0.3/W$ (d),
    $A_R=0.2/W$ (e) and $A_R=0.1/W$ (f) and panels g, h, i show
    $I_{q,c}(\Phi=0,eV/\Delta)$ and $I_{q,c}(\Phi=\pi,eV/\Delta)$ for
    $A_I=0.1/W$ and $A_R=0.3/W$ (g), $A_R=0.2/W$ (h) and $A_R=0.1/W$
    (i). We use $\Gamma/\Delta=1$, $\Gamma'/\Delta=0$, $B_R=B_I=0$,
    and $\epsilon_x=\epsilon_y=0$. The notation $W$ is used for the
    band-width.
  \label{fig:figureA-nonlog}
  }
\end{figure*}

\begin{figure*}[htb]
  \includegraphics[width=\textwidth]{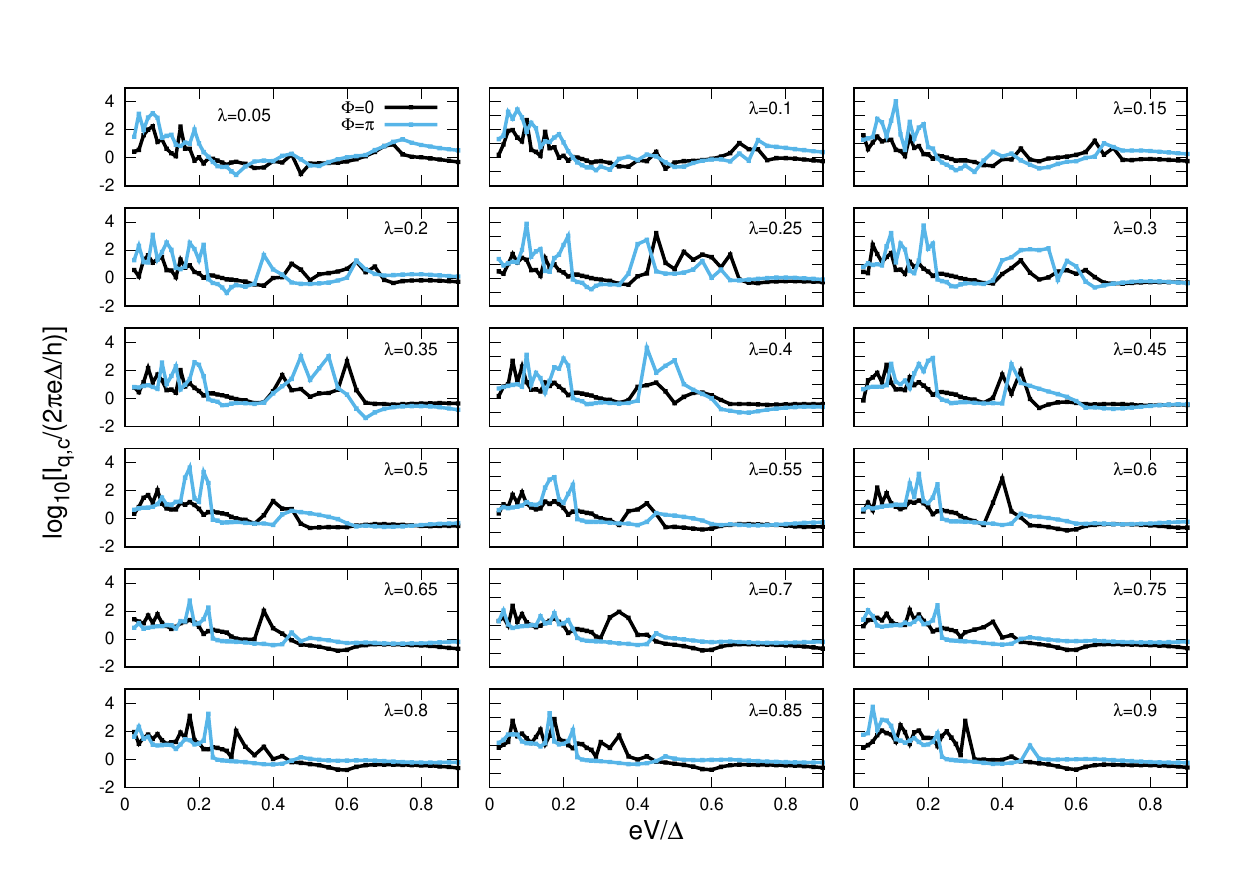}
  \caption{{\it The $eV/\Delta$-dependence of the quartet critical
      current.} The figure shows the reduced
    voltage-$eV/\Delta$-dependence of the logarithm of the quartet
    critical currents
    $\log_{10}\left(I_{q,c}(\Phi=0,eV/\Delta)\right)$ and
    $\log_{10}\left(I_{q,c}(\Phi=\pi,eV/\Delta)\right)$ at $\Phi=0$
    and $\Phi=\pi$ respectively. The panels show increasing values of
    $\lambda=\Gamma'/\Gamma$, where $\Gamma$ is the ``direct''
    coupling between $D_x$ and $S_a$, $S_{c,1}$ and between $D_y$ and
    $S_b$, $S_{c,2}$. The crossed coupling $\Gamma'$ is in between
    $D_x$ and $S_b$, $S_{c,2}$ and between $D_y$ and $S_a$,
    $S_{c,1}$. We used $\Gamma/\Delta=1$, $A_R=0.3/W$,
    $A_I=0.05/W$, and $B_R=B_I=0$, where $W$ is the
    band-width. The on-site energies $\epsilon_x=\epsilon_y=0$ are
    vanishingly small.
  \label{fig:toto1} 
  }
\end{figure*}

\begin{figure}[htb]
  \includegraphics[bb=60bp 52bp 190bp 200bp,clip,width=.85\columnwidth]{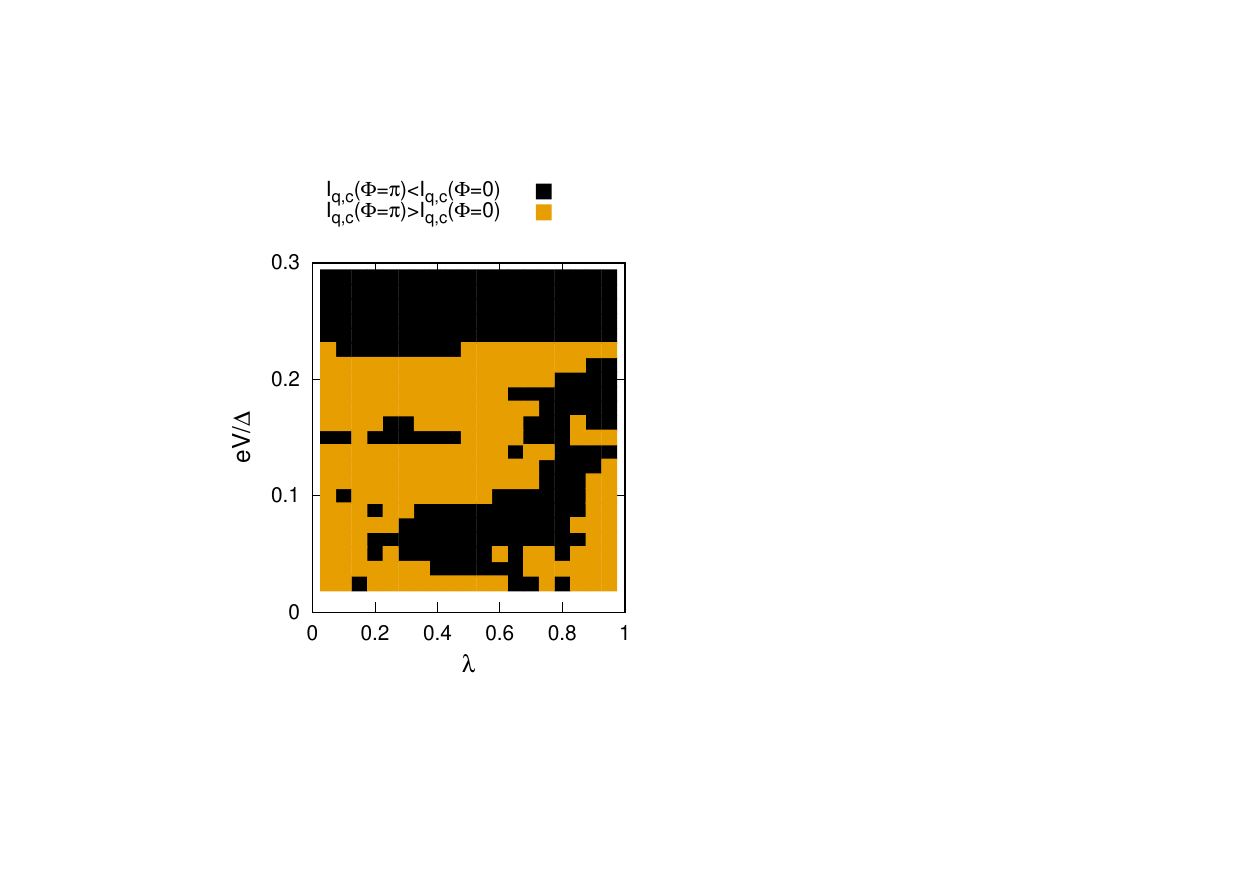}
  \caption{{\it The sign of the inversion.} The figure shows the sign
    of $I_{q,c}(\lambda,\Phi=\pi,eV/\Delta) -
    I_{q,c}(\lambda,\Phi=0,eV/\Delta)$, where
    $\lambda=\Gamma'/\Gamma$. Inversion corresponds to
    $I_{q,c}(\lambda,\Phi=\pi,eV/\Delta) -
    I_{q,c}(\lambda,\Phi=0,eV/\Delta)>0$.  The parameter $\Gamma$ is
    the ``direct'' coupling between $D_x$ and $(S_a,S_{c,1})$ or
    between $D_y$ and $(S_b,S_{c,2})$. The parameter $\Gamma'$ is in
    between $D_x$ and $(S_b,S_{c,2})$ or $D_y$ and $(S_a,S_{c,1})$. We
    used $\Gamma/\Delta=1$, $A_R=0.3/W$, $A_I=0.05/W$, and
    $B_R=B_I=0$, where $W$ is the band-width. The on-site
    energies $\epsilon_x=\epsilon_y=0$ are vanishingly small.
    \label{fig:toto2}
  }
\end{figure}

\begin{figure*}[htb]
  \begin{minipage}{.7\textwidth}
    \includegraphics[bb=62bp 52bp 270bp 200bp,clip,width=.8\textwidth]{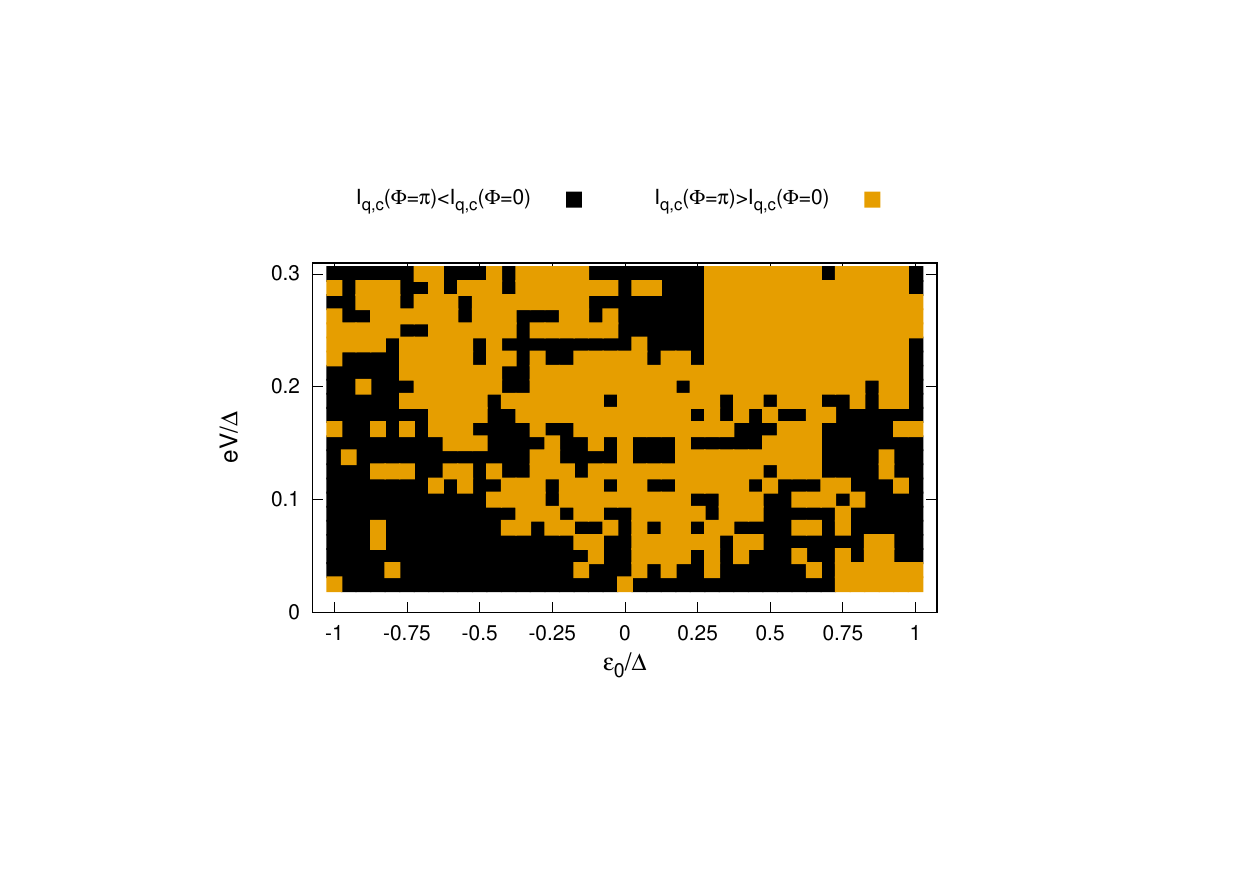}
    \end{minipage}\begin{minipage}{.28\textwidth}
  \caption{{\it The sign of the inversion.} The figure is similar to
    the previous figure~\ref{fig:toto2}, but now the sign of
    $I_{q,c}(\epsilon_0/\Delta,\Phi=\pi,eV/\Delta) -
    I_{q,c}(\epsilon_0,\Phi=0,eV/\Delta)$ is shown in the
    $(\epsilon_0/\Delta,eV/\Delta)$ plane. We used $\Gamma/\Delta=1$,
    $A_R=0.3/W$, $A_I=0.05/W$, $B_R=B_I=0$ (where
    $W$ is the band-width) and $\lambda=\Gamma'/\Gamma=1/2$. The
    on-site energies are identical for both quantum dots:
    $\epsilon_x=\epsilon_y=\epsilon_0$.
  \label{fig:toto4}
  }
  \end{minipage}
\end{figure*}

Overall the transition between figures~\ref{fig:evolution}b
and~\ref{fig:evolution}c implies negative sign for the ``split
quartets'' transmitting a Cooper pair into $S_{c,1}$ and another one
into $S_{c,2}$, and a positive sign for the ``unsplit quartets'',
corresponding to two Cooper pairs transmitted into $S_{c,1}$ or into
$S_{c,2}$. The resulting opposite signs in the ``split'' and
``unsplit'' channels imply the inversion, see our previous paper~I
\cite{paperI}.  This physical picture for ``emergence of inversion in
the $V=0^+$ limit'' is confirmed by the calculations presented in
Appendix~\ref{sec:ex-double}.

Thus, we demonstrated that ``inversion between $\Phi=0$ and
$\Phi=\pi$'' or ``absence of inversion'' is linked to the quantum dot
single-particle wave-functions, i.e. to the number of nodes in their
wave-functions.

\section{Numerical results}
\label{sec:num-2nodes}

In this section, we present a selection of the numerical results for
the phenomenological model of double 0D quantum dot with normal lead
in series, see figure~\ref{fig:geometrie-4T-double-dot-remote} and the
Hamiltonian in subsection~\ref{sec:remote-0D}. The quantum dots $D_x$
and $D_y$ are coupled to each other by the tunneling amplitude
$\Sigma^{(0)}$ defined in Eq.~(\ref{eq:H-xy}). The coupling
$\Gamma=\left(\Sigma^{(2)}\right)^2/W$ is defined for hopping between
$D_x$ and $S_a$, $D_x$ and $S_{c,1}$, $D_y$ and $S_b$, $D_y$ and
$S_{c,2}$, see $\Sigma^{(2)}$ in Eq.~(\ref{eq:Sigma(2)}). We add the
coupling $\Gamma'=\left(\Sigma'^{(2)}\right)^2/W$ between $D_x$ and
$S_b$, $D_x$ and $S_{c,2}$, $D_y$ and $S_a$, $D_y$ and $S_{c,1}$, see
$\Sigma'^{(2)}$ in Eq.~(\ref{eq:Sigma-prime-2}). The Green's functions
of $N'$ are given by
\begin{eqnarray}
  \label{eq:g-1}
  \hat{g}_{N'_x,N'_y}^{1,1}=\hat{g}_{N'_y,N'_x}^{1,1}=A_R +
  i
  A_I\\ \hat{g}_{N'_x,N'_y}^{2,2}=\hat{g}_{N'_y,N'_x}^{2,2}=-A_R
  + i
  A_I\\ \hat{g}_{N'_x,N'_x}^{1,1}=\hat{g}_{N'_y,N'_y}^{1,1}=B_R
  + i
  B_I\\ \hat{g}_{N'_x,N'_x}^{2,2}=\hat{g}_{N'_y,N'_y}^{2,2}=-B_R
  + i B_I ,
  \label{eq:g-2}
\end{eqnarray}
where $(A_R,\, A_I)$, and $(B_R,\, B_I)$ are four real-valued
parameters for the nonlocal and local Green's functions
$\hat{g}_{N'_x,N'_y} = \hat{g}_{N'_y,N'_x}$ and $\hat{g}_{N'_x,N'_x}$,
$\hat{g}_{N'_y,N'_y}$ respectively, within the assumption
$\hat{g}_{N'_x,N'_x} = \hat{g}_{N'_y,N'_y}$. {All of
  the numerical calculations presented below are within the assumption
  $B_R=B_I=0$, which is justified as mimicking perfect transmission
  for highly transparent extended interfaces between the
  superconductors and the ballistic 2D metal. This assumption yields
  vanishingly small value for the local Green's function involving
  ``U-turn'' of the quasiparticle coming back on the interface, a
  process that is not possible at high transparency in the absence of
  disorder. In addition to being physically motivated, the condition
  $B_R=B_I=0$ allows restricting the parameter space to be explored in
  the numerical calculations.}
 
The assumption of 0D quantum dots implies that the quartet critical
current depends on the value of the Green's function crossing the
conductor $N'$ between $N'_x$ and $N'_y$, see $N'$, $N'_x$ and $N'_y$
in figure~\ref{fig:geometrie-4T-double-dot-remote}. The nonlocal
Green's function crossing $N'$ oscillates with the combinations
$\cos(k_F R_0)$ and $\sin(k_F R_0)$, where $R_0$ is the separation
between $N'_x$ and $N'_y$, and $k_F$ is the Fermi wave-vector. Sharp
resonance peaks emerge at fixed $k_F R_0$ in the voltage-$V$
dependence of the quartet current, see the previous
section~\ref{sec:mechanism} and the numerical results in the present
section. The interplay between those $k_F R_0$-oscillations in space
and the sharp Floquet resonances in the voltage-dependence of the
quartet critical current is expected {to produce
  log-normal distribution of strong sample-to-sample fluctuations,}
where different samples are characterized by different values of
$R_0$. In this section, we specifically investigate the regime of
strong log-normal distribution of sample-to-sample fluctuations and
make the physically-motivated assumption that fixing $A_R$, $A_I$
yields quartet critical current-voltage dependence that is
qualitatively representative of the signal in this regime of strong
sample-to-sample fluctuations.

The codes have been developed over the last few years
\cite{HGE,Sotto,FWS,engineering,paperII,long-distance,Heiblum,Berry}. They
are based on recursive calculations as a function of the harmonics of
the Josephson frequency \cite{Cuevas,AVERIN} (see also the Appendix of
Ref.~\onlinecite{Sotto}) and on sparse matrix algorithms for matrix
products. We specifically adapted the code of our recent
Ref.~\onlinecite{long-distance} to include the local and nonlocal
Green's functions given by Eqs.~(\ref{eq:g-1})-(\ref{eq:g-2}).

In the forthcoming figures~\ref{fig:figureA-nonlog}
and~\ref{fig:toto1}, we present numerical results for the
voltage-dependence of the quartet critical current $I_{q,c}$ defined
as
\begin{eqnarray}
  \label{eq:Iqc-AA}
&&
  I_{q,c}(\Phi=0,eV/\Delta)=\\ &&\mbox{max}_{\varphi_q}\left[I_q(\Phi=0,\varphi_q,eV/\Delta)
    -I_q(\Phi=0,-\varphi_q,eV/\Delta)\right]\nonumber\\\nonumber
  &-&\mbox{min}_{\varphi_q}\left[I_q(\Phi=0,\varphi_q,eV/\Delta)-I_q(\Phi=0,-\varphi_q,eV/\Delta)\right]\\ \label{eq:Iqc-BB}
  && I_{q,c}(\Phi=\pi,eV/\Delta)=\\\nonumber &&
  \mbox{max}_{\varphi_q}\left[I_q(\Phi=\pi,\varphi_q,eV/\Delta)-I_q(\Phi=\pi,-\varphi_q,eV/\Delta)\right]\\ &-&
  \mbox{min}_{\varphi_q}\left[I_q(\Phi=\pi,\varphi_q,eV/\Delta)-I_q(\Phi=\pi,-\varphi_q,eV/\Delta)\right]
  .\nonumber
\end{eqnarray}
In addition, we present colormaps for the sign of $I_{q,c}$ as a
function of the model parameters (see the forthcoming
figures~\ref{fig:toto2} and~\ref{fig:toto4}).

Figure~\ref{fig:figureA-nonlog} shows the normalized bias
voltage-$eV/\Delta$-dependence of the quartet critical currents
$I_{q,c}(\Phi=0,eV/\Delta)$ and $I_{q,c}(\Phi=\pi,eV/\Delta)$ at the
flux values $\Phi=0$ and $\Phi=\pi$. The coupling parameters
$\Gamma/\Delta=1$ and $\Gamma'/\Delta=0$ are used in
figure~\ref{fig:figureA-nonlog}. 

Figures~\ref{fig:figureA-nonlog}a, \ref{fig:figureA-nonlog}b and
\ref{fig:figureA-nonlog}c correspond to absence of relaxation with
$A_I=0$, producing small quartet critical current.
Figures~\ref{fig:figureA-nonlog}d, \ref{fig:figureA-nonlog}e,
\ref{fig:figureA-nonlog}f and~\ref{fig:figureA-nonlog}g,
\ref{fig:figureA-nonlog}h, \ref{fig:figureA-nonlog}i correspond to the
nonvanishingly small relaxation parameters $A_I=0.05/W$ and
$A_I=0.1/W$ respectively, producing large quartet critical current at
small voltage ratio $eV/\Delta$. In addition, sharp resonances emerge
in the variations of the quartet critical current as a function of
$eV/\Delta$ for $A_I=0.05/W$ and $A_I=0.1/W$ in
figures~\ref{fig:figureA-nonlog}d, \ref{fig:figureA-nonlog}e,
\ref{fig:figureA-nonlog}f and~\ref{fig:figureA-nonlog}g,
\ref{fig:figureA-nonlog}h, \ref{fig:figureA-nonlog}i
respectively. The numerical results in
figure~\ref{fig:figureA-nonlog} are in agreement with the general
mechanism of section~\ref{sec:mechanism}.

Figure~\ref{fig:toto1} shows the normalized
voltage-$eV/\Delta$-dependence of the quartet critical current
$I_{q,c}$, in semi-logarithmic scale, now with nonvanishingly small
$\lambda= \Gamma'/\Gamma$, and with $A_I=0.05/W$. Those sharp peaks in
the quartet critical current are expected to change as the value of
the normal metal Green's functions is varied in
Eqs.~(\ref{eq:g-1})-(\ref{eq:g-2}). The quartet critical current
plotted as a function of the bias voltage $V$ is normal in
log-scale. The corresponding peaks are expected to evolve as $k_F R_0$
is varied and thus, sample-to-sample fluctuations are expected to have
log-normal distribution.

In addition, multiple cross-overs are obtained in
figure~\ref{fig:toto1} between
$I_{q,c}(\Phi=\pi,eV/\Delta)-I_{q,c}(\Phi=0,eV/\Delta)>0$
(i.e. inversion) and
$I_{q,c}(\Phi=\pi,eV/\Delta)-I_{q,c}(\Phi=0,eV/\Delta)<0$
(i.e. noninverted behavior) in a low-voltage window.

Figure~\ref{fig:toto2} shows the sign of
$I_{q,c}(\lambda,\Phi=\pi,eV/\Delta) -
I_{q,c}(\lambda,\Phi=0,eV/\Delta)$ as a function of the parameters
$\lambda=\Gamma'/\Gamma$ (on the $x$-axis) and $eV/\Delta$ (on the
$y$-axis). Inversion $I_{q,c}(\lambda,\Phi=\pi,eV/\Delta) -
I_{q,c}(\lambda,\Phi=0,eV/\Delta)>0$ is obtained at low $eV/\Delta$
for $\lambda=0$, see the previous section~\ref{sec:mecha-inversion}
and Appendix~\ref{app:0+} for a discussion of the inversion in the
$eV/\Delta=0^+$ limit. In figure~\ref{fig:toto2}, increasing the
coupling $\Gamma'$ between $D_x$ and $(S_b,S_{c,2})$ or between $D_y$
and $(S_a,S_{c,1})$ [in addition to $\Gamma$ between $D_x$ and
  $(S_a,S_{c,1})$ or between $D_y$ and $(S_b,S_{c,2})$], makes the
double 0D quantum dot behave closer to a pair of single quantum
dots. This favors the ``noninverted behavior'' typical of single
quantum dots, as opposed to the ``inverted behavior'' appearing at
$\lambda=\Gamma'/\Gamma=0$ in a double 0D quantum dot, see
Appendix~\ref{app:0+}.

The values $\lambda=\Gamma'/\Gamma$ such that $0.2\alt \lambda \alt
0.6$ in figure~\ref{fig:toto2} generically yield ``inversion'' at low
bias-voltage $V$, followed by ``absence of inversion'' at higher
$V$-values. Now, we show that ``absence of inversion'' can appear at
low bias-voltage $V$ {if the on-site energies
  $\epsilon_x=\epsilon_y\equiv \epsilon_0$ in the double quantum dot
  Hamiltonian given by Eqs.~(\ref{eq:H-x-0D})-(\ref{eq:H-y-0D})
  produce detuning from perfectly opposite levels.}  Namely, the
parameter $\lambda=\Gamma'/\Gamma$ is set to $\lambda=1/2$ in
figure~\ref{fig:toto4}. Moving away from $\epsilon_0/\Delta=0$ in
figure~\ref{fig:toto4} produces typical voltage-dependence of
$I_{q,c}(\lambda,\Phi=\pi,eV/\Delta) -
I_{q,c}(\lambda,\Phi=0,eV/\Delta)$ with ``noninverted behavior'' at
low $V<V_*$ and ``inversion'' at higher $V>V_*$, see for instance the
moderately small values $-0.5 \alt \epsilon_0/\Delta \alt 0.5$ in
figure~\ref{fig:toto4}. With the considered parameters, the ratio
$eV_*/\Delta$ is in the range $eV_*/\Delta\approx 0.1$. Thus, our
model is compatible with the experimental data \cite{HGE} shown in
figure~\ref{fig:experiment}b.

\section{Low-voltage limit}
\label{sec:low-V}

{In this section, we focus on the low-voltage limit
  and insert in Eq.~(\ref{eq:I-N-nu3}) the equilibrium fully dressed
  advanced and retarded Green's functions. We present an explanation
  for why the numerical calculations of the preceding
  section~\ref{sec:num-2nodes} suggest that the line-width broadening
  $\delta_l$ in Eq.~(\ref{eq:delta_l}) appears to be much smaller than
  $\sim A_I$ in Eqs.~(\ref{eq:g-1})-(\ref{eq:g-2}).}

{As in the previous numerical calculations presented
  in section~\ref{sec:num-2nodes}, we consider normal leads in series,
  see figure~\ref{fig:geometrie-4T-double-dot-remote} and the
  Hamiltonian in subsection~\ref{sec:remote-0D}. The Dyson
  Eqs.~(\ref{eq:GDD-1}) take the form
\begin{eqnarray}
  \label{eq:DYSONA}
&&  \left((\omega-i\eta) -\hat{\Gamma}_{x,x}^{(a,a)} - \hat{\Gamma}_{x,x}^{(c_1,c_1)}
  -\hat{\Gamma}_{x,x}^{(N'_x,N'_x)}\right) \hat{G}_{x,x}\\
  \nonumber
&&-\hat{\Gamma}_{x,y}^{(N'_x,N'_y)} \hat{G}_{y,x}=\hat{g}_{x,x}\\
\label{eq:DYSONB}
&&\left((\omega-i\eta) -\hat{\Gamma}_{y,y}^{(b,b)} - \hat{\Gamma}_{y,y}^{(c_1,c_1)}
-\hat{\Gamma}_{y,y}^{(N'_y,N'_y)}\right) \hat{G}_{y,x}\\
&&-\hat{\Gamma}_{y,x}^{(N'_y,N'_x)} \hat{G}_{x,x}=0
,\nonumber
\end{eqnarray}
with
\begin{eqnarray}
  \label{eq:H-def-A}
  \hat{\Gamma}_{x,x}^{(a,a)}&=&-\Gamma_a \left(\begin{array}{cc}
    0 & \exp(i\varphi_a)\\
    \exp(-i\varphi_a) & 0 \end{array}\right)\\
  \hat{\Gamma}_{x,x}^{(c_1,c_1)}&=&-\Gamma_{c_1} \left(\begin{array}{cc}
    0 & \exp(i\varphi_{c,1})\\
    \exp(-i\varphi_{c,1}) & 0 \end{array}\right)\\
  \hat{\Gamma}_{y,y}^{(b,b)}&=&-\Gamma_b \left(\begin{array}{cc}
    0 & \exp(i\varphi_b)\\
    \exp(-i\varphi_b) & 0 \end{array}\right)\\
  \hat{\Gamma}_{y,y}^{(c_2,c_2)}&=&-\Gamma_{c_2} \left(\begin{array}{cc}
    0 & \exp(i\varphi_{c,2})\\
    \exp(-i\varphi_{c,2}) & 0 \end{array}\right)
  ,
\end{eqnarray}
where the couplings $\Gamma_a$, $\Gamma_b$, $\Gamma_{c,1}$ and
$\Gamma_{c,2}$ are between the double quantum dot and the
superconducting leads $S_a$, $S_b$, $S_{c,1}$ and $S_{c,2}$, see
figure~\ref{fig:geometrie-4T-double-dot-remote} and the analogous
Eqs.~(\ref{eq:Gamma-a})-(\ref{eq:Gamma-c2}). In addition,
$\hat{\Gamma}_{x,x}^{(N'_x,N'_x)}$,
$\hat{\Gamma}_{x,y}^{(N'_x,N'_y)}$,
$\hat{\Gamma}_{y,y}^{(N'_y,N'_y)}$ and
$\hat{\Gamma}_{y,x}^{(N'_y,N'_x)}$ are given by
\begin{eqnarray}
  \label{eq:Gamma-non-loc-A}
  \hat{\Gamma}_{x,x}^{(N'_x,N'_x)}=\Gamma_{N'_x,N'_x}
  \left(\begin{array}{cc} Wg_{N'_x,N'_x}^{1,1} & 0\\
    0 & W g_{N'_x,N'_x}^{2,2} \end{array}\right)\\
\hat{\Gamma}_{x,y}^{(N'_x,N'_y)}=\Gamma_{N'_x,N'_y}
  \left(\begin{array}{cc} Wg_{N'_x,N'_y}^{1,1} & 0\\
    0 & W g_{N'_x,N'_y}^{2,2} \end{array}\right)\\
  \hat{\Gamma}_{y,y}^{(N'_y,N'_y)}=\Gamma_{N'_y,N'_y}
  \left(\begin{array}{cc} Wg_{N'_y,N'_y}^{1,1} & 0\\
    0 & W g_{N'_y,N'_y}^{2,2} \end{array}\right)\\
\hat{\Gamma}_{y,x}^{(N'_y,N'_x)}=\Gamma_{N'_y,N'_x}
  \left(\begin{array}{cc} Wg_{N'_y,N'_x}^{1,1} & 0\\
    0 & W g_{N'_y,N'_x}^{2,2} \end{array}\right)
  ,
\label{eq:Gamma-non-loc-B}
\end{eqnarray}
with $\Gamma_{N'_x,N'_x}=\left(\Sigma_{x,N'_x}\right)^2/W$,
$\Gamma_{N'_x,N'_y}=\Sigma_{x,N'_x}\Sigma_{y,N'_y}/W$,
$\Gamma_{N'_y,N'_x}=\Sigma_{x,N'_y}\Sigma_{x,N'_x}/W$,
$\Gamma_{N'_y,N'_y}=\left(\Sigma_{x,N'_y}\right)^2/W$, where
$N'_x$ and $N'_y$ are the normal-metal tight-binding sites making
the contacts with the 0D quantum dots $D_x$ and $D_y$ respectively.}
\begin{figure*}[htb]
    \includegraphics[width=.49\textwidth]{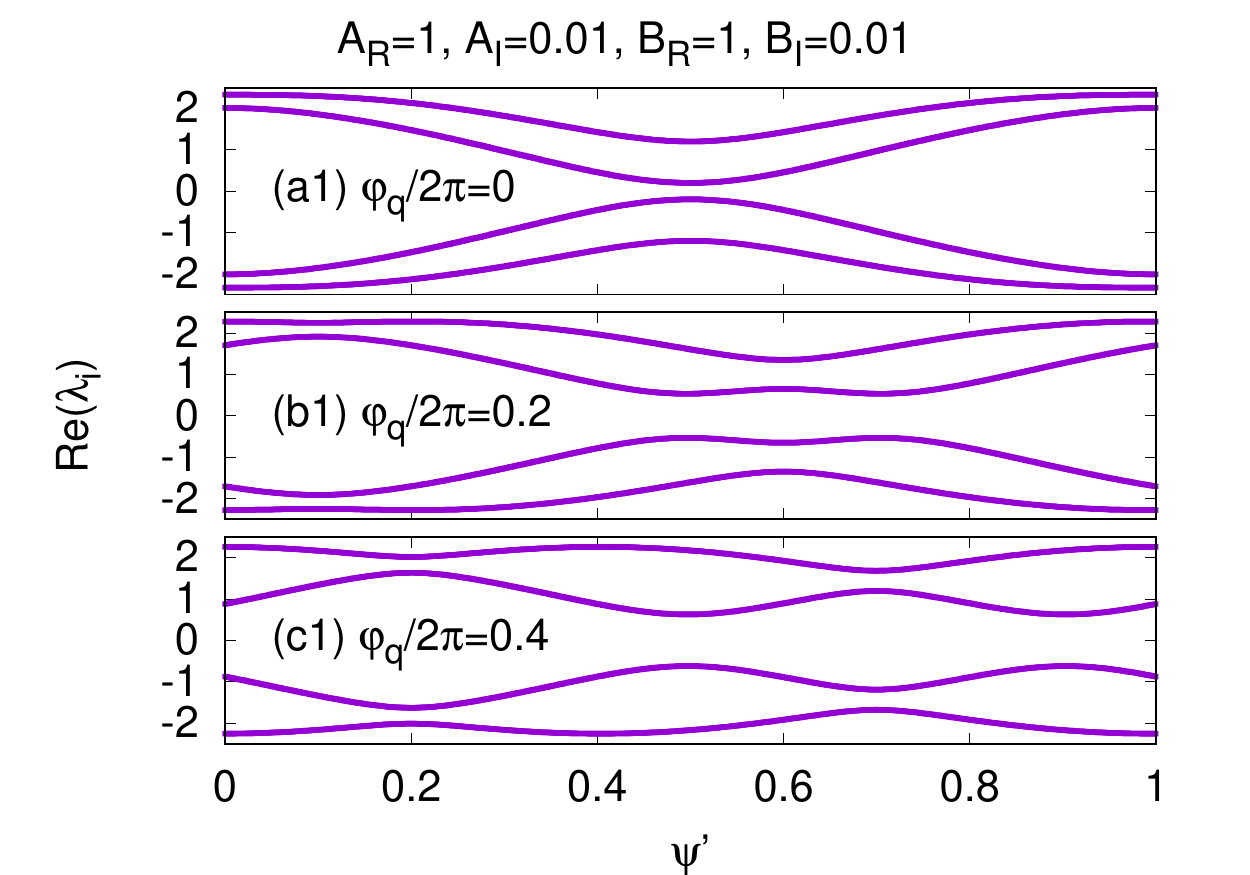}\includegraphics[width=.49\textwidth]{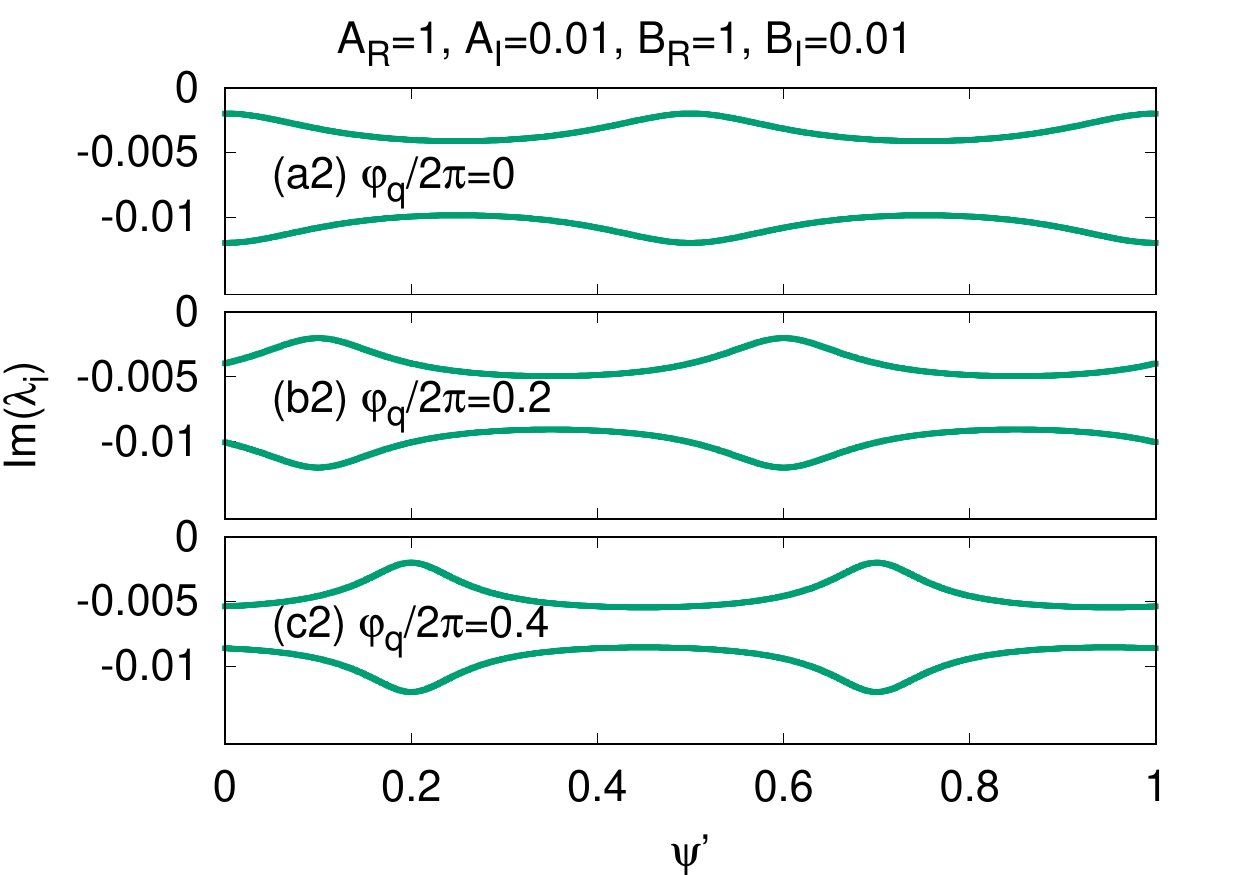}
    \caption{{{\it The real and imaginary parts of the
          ABS energies in the infinite-gap limit:} Panels (a1)-(c1)
        and (a2)-(c2) show the real and imaginary parts of the ABS
        energies respectively, for the $4\times 4$ Hamiltonian given
        by Eq.~(\ref{eq:H-def-A})-(\ref{eq:H-infini-gene}), plotted as
        a function of the variable $\psi'$ defined as
        $\varphi_a=\varphi_a^{(0)}+\psi'$,
        $\varphi_b=\varphi_b^{(0)}-\psi'$
        $\varphi_{c,1}=\varphi_{c}^{(0)}$ and
        $\varphi_{c,2}=\varphi_{c}^{(0)}$, with
        $\varphi_q=\varphi_a^{(0)}+\varphi_b^{(0)}-2\varphi_c^{(0)}$. The
        parameters take the general values indicated on the figure,
        with in addition
        $\Gamma_a=\Gamma_b=\Gamma_{c,1}=\Gamma_{c,2}=1$ taken as the
        unit of energy for the coupling to the superconducting leads,
        and $\Gamma_{loc}=0.7$, $\Gamma_{non\,loc}=0.5$ for the
        ``local'' and ``nonlocal'' electron-electron channels, see
        Eqs.~(\ref{eq:Gamma-non-loc-A})-(\ref{eq:Gamma-non-loc-B}).
        Panels a2, b2 and c2 reveal that $|\mbox{Im}(\lambda_i)|$ can
        be small compared to $A_I$ or $B_I$ (but not orders of
        magnitude smaller), which implies small $\delta_l$ in
        Eq.~(\ref{eq:delta_l}), and compatibility with the large
        quartet current reported in the numerical results presented in
        the above section~\ref{sec:num-2nodes}.}
  \label{fig:toto5}
  }
\end{figure*}

{The Dyson Eqs.~(\ref{eq:DYSONA})-(\ref{eq:DYSONB})
  define the infinite-gap Hamiltonian
  \begin{widetext}
  \begin{equation}
    \label{eq:H-infini-gene}
    {\cal H}^{(\infty)}=
    \left(\begin{array}{cc}
      \hat{\Gamma}_{x,x}^{(a,a)} +\hat{\Gamma}_{x,x}^{(c_1,c_1)}+\hat{\Gamma}_{x,x}^{(N'_x,N'_x)} &
      \hat{\Gamma}_{x,y}^{(N'_x,N'_y)}\\
      \hat{\Gamma}_{y,x}^{(N'_y,N'_x)} &
      \hat{\Gamma}_{y,y}^{(b,b)} +\hat{\Gamma}_{y,y}^{(c_2,c_2)}+\hat{\Gamma}_{y,y}^{(N'_y,N'_y)}
    \end{array}\right)
    ,
  \end{equation}
  \end{widetext}
which is now considered in absence of the local couplings
$\hat{\Gamma}_{x,x}^{(N'_x,N'_x)}=\hat{\Gamma}_{y,y}^{(N'_y,N'_y)}\equiv
0$, corresponding to the use of $B_R=B_I=0$ in the above numerical
calculations, see the discussion following
Eqs.~(\ref{eq:g-1})-(\ref{eq:g-2}) in section
~\ref{sec:num-2nodes}. Then, the square of the $4\times 4$
infinite-gap Hamiltonian decouples into a pair of $2\times 2$ blocks,
see also Appendix~\ref{app:0+}:
\begin{equation}
  \left({\cal H}^{(\infty)}\right)^2=
  \left(\begin{array}{cccc} \epsilon_{a,c_1}^{1,1} & 0 & 0 & \kappa_{1,2}\\
    0 & \epsilon_{a,c_1}^{2,2} & \kappa_{2,1} & 0\\
    0 & \kappa'_{1,2} & \epsilon_{b,c_2}^{1,1} & 0\\
    \kappa'_{2,1} & 0 & 0 & \epsilon_{b,c_2}^{2,2}
  \end{array} \right)
  ,
\end{equation}
with
\begin{eqnarray}
  \epsilon_{a,c_1}^{1,1} &=&|\gamma_{D_x,D_x}|^2 + \left(\Gamma_{N'_x,N'_y}\right)^2 W^2 g_{N'_x,N'_y}^{1,1} g_{N'_y,N'_x}^{1,1}\\
  \epsilon_{a,c_1}^{2,2} &=&|\gamma_{D_x,D_x}|^2 + \left(\Gamma_{N'_x,N'_y}\right)^2 W^2 g_{N'_x,N'_y}^{2,2} g_{N'_y,N'_x}^{2,2}\\
  \epsilon_{b,c_2}^{1,1} &=&|\gamma_{D_y,D_y}|^2 + \left(\Gamma_{N'_x,N'_y}\right)^2 W^2 g_{N'_x,N'_y}^{1,1} g_{N'_y,N'_x}^{1,1}\\
  \epsilon_{b,c_2}^{2,2} &=&|\gamma_{D_y,D_y}|^2 + \left(\Gamma_{N'_x,N'_y}\right)^2 W^2 g_{N'_x,N'_y}^{2,2} g_{N'_y,N'_x}^{2,2}\\
\kappa_{1,2}&=& \Gamma_{N'_x,N'_y} W\left(g_{N'_x,N'_y}^{2,2} \gamma_{D_x,D_x} + g_{N'_x,N'_y}^{1,1} \gamma_{D_y,D_y}\right)\\
\kappa_{2,1}&=& \Gamma_{N'_x,N'_y} W\left(g_{N'_x,N'_y}^{1,1} \overline{\gamma_{D_x,D_x}} + g_{N'_x,N'_y}^{2,2} \overline{\gamma_{D_y,D_y}}\right)\\
\kappa'_{1,2}&=& \Gamma_{N'_x,N'_y} W\left(g_{N'_y,N'_x}^{1,1} \gamma_{D_x,D_x} + g_{N'_y,N'_x}^{2,2} \gamma_{D_y,D_y}\right)\\
\kappa'_{2,1}&=& \Gamma_{N'_x,N'_y} W\left(g_{N'_y,N'_x}^{2,2} \overline{\gamma_{D_x,D_x}} + g_{N'_y,N'_x}^{1,1} \overline{\gamma_{D_y,D_y}}\right)
,
\end{eqnarray}
where
\begin{eqnarray}
  \label{eq:gamma-Dx-Dx}
\gamma_{D_x,D_x}&=&\Gamma_a \exp(i\varphi_a)+\Gamma_{c_1}
\exp(i\varphi_{c,1})\\
\label{eq:gamma-Dy-Dy}
\gamma_{D_y,D_y}&=&\Gamma_b
\exp(i\varphi_b)+\Gamma_{c_2} \exp(i\varphi_{c,2}).
\end{eqnarray}
Within each
$2\times 2$ block, the eigenvalues take the form
\begin{equation}
  \lambda_{\pm}=\frac{1}{2}\left[
    \epsilon_{a,c_1}^{1,1}+\epsilon_{b,c_2}^{2,2}\pm
    \sqrt{\left(\epsilon_{a,c_1}^{1,1}-\epsilon_{b,c_2}^{2,2}\right)^2
      +4 \kappa_{1,2} \kappa'_{2,1}}\right]
,
\end{equation}
where the following quantities
\begin{eqnarray}
  \label{eq:real-valued1}
  &&  \epsilon_{a,c_1}^{1,1}+\epsilon_{b,c_2}^{2,2}=\\
  \nonumber
&&  |\gamma_{D_x,D_x}|^2 + |\gamma_{D_y,D_y}|^2
  + 2 \left(\Gamma_{N'_x,N'_y}\right)^2 W^2\left(A_R^2-A_I^2\right)\\
  \label{eq:real-valued2}
 && \left(\epsilon_{a,c_1}^{1,1}-\epsilon_{b,c_2}^{2,2}\right)^2
  +4 \kappa_{1,2} \kappa'_{2,1}=\\
  \nonumber
&& \left(|\gamma_{D_x,D_x}|^2 - |\gamma_{D_y,D_y}|^2\right)^2
  -16 \left(\Gamma_{N'_x,N'_y}\right)^2 W^2 A_R^2 A_I^2\\
  \nonumber
  &&+4 \left(\Gamma_{N'_x,N'_y}\right)^2 W^2 A_R^2 \left|\gamma_{D_x,D_x} - \gamma_{D_y,D_y}\right|^2\\
  \nonumber
&&-4 \left(\Gamma_{N'_x,N'_y}\right)^2 W^2 A_I^2 \left|\gamma_{D_x,D_x} + \gamma_{D_y,D_y}\right|^2
\end{eqnarray}
are both real-valued.}

{It can also be shown that $\lambda_\pm$ is
  real-valued and positive at small $A_I$ because
\begin{eqnarray}
  \left|\gamma_{D_x,D_x} \overline{\gamma_{D_y,D_y}}
  - \left(\Gamma_{N'_x,N'_y}\right)^2 W^2 g_{N'_x,N'_y}^{1,1}g_{N'_y,N'_x}^{2,2} \right|^2
\end{eqnarray}
is positive, which implies
\begin{eqnarray}
  \nonumber
&&|\gamma_{D_x,D_x}|^2 |\gamma_{D_y,D_y}|^2
+
\left(\Gamma_{N'_x,N'_y}\right)^4 W^4
g_{N'_x,N'_y}^{1,1}g_{N'_y,N'_x}^{1,1}g_{N'_x,N'_y}^{2,2}g_{N'_y,N'_x}^{2,2}\\
  \nonumber
  &>&
  \left(\Gamma_{N'_x,N'_y}\right)^2 W^2 g_{N'_x,N'_y}^{1,1}g_{N'_y,N'_x}^{2,2}
  \gamma_{D_y,D_y} \overline{\gamma_{D_x,D_x}}\\
&&  +
  \left(\Gamma_{N'_x,N'_y}\right)^2 W^2 g_{N'_x,N'_y}^{2,2}g_{N'_y,N'_x}^{1,1}
  \gamma_{D_x,D_x} \overline{\gamma_{D_y,D_y}}
  ,
\end{eqnarray}
where we used the approximation
$\overline{g_{N'_x,N'_y}^{1,1}}\simeq g_{N'_x,N'_y}^{1,1}$ and
$\overline{g_{N'_x,N'_y}^{2,2}}\simeq g_{N'_x,N'_y}^{2,2}$ at small
$A_I$. Thus, we obtain
\begin{eqnarray}
  &&
  \left(|\gamma_{D_x,D_x}|^2 + \left(\Gamma_{N'_x,N'_y}\right)^2 W^2g_{N'_x,N'_y}^{1,1}g_{N'_y,N'_x}^{1,1}
  \right)\\
  \nonumber
&&\times  \left(|\gamma_{D_y,D_y}|^2 + \left(\Gamma_{N'_x,N'_y}\right)^2 W^2g_{N'_x,N'_y}^{2,2}g_{N'_y,N'_x}^{2,2}
  \right)\\
  \nonumber
  &>&
  \left(\Gamma_{N'_x,N'_y}\right)^2W^2
  \left(g_{N'_x,N'_y}^{2,2} \gamma_{D_x,D_x} + g_{N'_x,N'_y}^{1,1}\gamma_{D_y,D_y}\right)\\
  \nonumber
  &&\times
  \left(g_{N'_y,N'_x}^{2,2} \overline{\gamma_{D_x,D_x}} + g_{N'_y,N'_x}^{1,1}\overline{\gamma_{D_y,D_y}}\right)
  ,
\end{eqnarray}
from what we deduce
\begin{equation}
  \label{eq:positive}
  \left(\epsilon_{a,c_1}^{1,1}+\epsilon_{b,c_2}^{2,2}\right)^2 >
  \left(\epsilon_{a,c_1}^{1,1}-\epsilon_{b,c_2}^{2,2}\right)^2
  + 4 \kappa_{1,2} \kappa'_{2,1}
\end{equation}
at small $A_I$.  Then, the eigenvalues $\lambda_{\pm}$ of the squared
infinite-gap Hamiltonian are both real-valued at arbitrary $A_I$ [see
  Eqs.~(\ref{eq:real-valued1})-(\ref{eq:real-valued2})] and positive
at small $A_I$ [see Eq.~(\ref{eq:positive})].}

{In addition, we carried out complementary numerical
  calculation for the eigenvalues of the equilibrium $4\times 4$
  infinite-gap Hamiltonian defined by
  Eqs.~(\ref{eq:H-def-A})-(\ref{eq:H-infini-gene}), see
  figure~\ref{fig:toto2} We find the possibility of imaginary part of
  the ABS energies smaller (but not orders of magnitude smaller) than
  $A_I$ and $B_I$ with general values of the parameters, i.e. not
  within the above-considered $B_R=B_I=0$ and small $A_I$. Thus, small
  $\delta_l$ has to be used in Eq.~(\ref{eq:delta_l}) at low bias
  voltage, which is compatible with the large quartet current in
  figure~\ref{fig:toto1}, similarly to the thermal noise of a
  superconducting weak link \cite{Yeyati-noise-equilibrium}.}

\section{Conclusions}
\label{sec:conclusions}

{Now, we provide final remarks.} The paper is
summarized in subsection~\ref{sec:summary}, and perspectives are
presented in subsection~\ref{sec:discussion}.

\subsection{Summary of the paper}
\label{sec:summary}

We proposed models of ballistic multiterminal Josephson junctions made
with superconducting leads evaporated on a 2D
metal. {The models consist of multilevel quantum dots
  connected to superconducting and normal leads.} We argued that the
qualitative physics of the two-Cooper pair resonance can be captured
with phenomenological double quantum dots connected to superconducting
and normal leads. The coupling to the nonproximitized regions of the
ballistic conductor produces relaxation, and a fraction of the
quartet-phase sensitive current is transmitted into the normal parts
of the circuit.

We found resonances in calculations that account for both the
time-periodic dynamics and small relaxation, if the multilevel or
double quantum dot support levels at opposite energies. A related
effect was previously found in the thermal noise of a superconducting
weak link \cite{Yeyati-noise-equilibrium}. {In
  addition, we addressed detuning from perfectly opposite energy
  levels.}

Noninverted-to-inverted cross-over numerically emerges as the bias
voltage increases, where ``inversion'' means ``larger quartet critical
current at half flux-quantum $\Phi=\pi$ than in zero field at
$\Phi=0$''. The corresponding cross-over voltage $V_*$ is small
compared to the superconducting gap, typically $eV_*\approx \Delta/10$
with the parameters of our calculations, which is compatible with the
recent Harvard group experiment \cite{HGE}.

\subsection{Perspectives}
\label{sec:discussion}

{A challenge for the theory is} to model devices that are quite
complex, with e.g. extended interfaces, four or more superconducting
leads, nonequilibrium voltage biasing conditions, loops connecting
superconducting terminals, possibly with radio-frequency
radiation. Direct diagonalizations of the Bogoliubov-de Gennes
Hamiltonian would apparently lead to prohibitive computational
expenses. In the field of mesoscopic superconductivity, Nazarov and
co-workers (see Ref.~\onlinecite{Nazarov-book} and references therein)
proposed and developed finite element theory for the
superconducting-normal metal circuits that describe the proximity
effect, i.e. the interplay between Andreev reflection and multiple
scattering on disorder. This dirty-limit circuit theory was proposed
for multiterminal Josephson junctions \cite{Padurariu,Padurariu2}.
The dirty limit implies short elastic mean free path, a condition that
is not directly met in the ballistic metals that are currently used in
some experiments on superconducting hybrid structures, such as carbon
nanotubes \cite{Pillet-CNT}, semiconducting nanowires
\cite{multiterminal-exp7} or graphene
\cite{Bretheau3,Park}. Specifically, tunneling spectroscopy
of carbon nanotube Josephson junctions \cite{Pillet-CNT} revealed
discrete ABS. Andreev molecules were realized with semiconducting
nanowires \cite{multiterminal-exp7}.  Evidence for superconducting
phase difference-sensitive continuum of ABS was obtained in
superconductor-graphene-superconductor Josephson junctions
\cite{Bretheau1,Bretheau2}. As it is mentioned above, microwave
experiments on short superconductor-graphene-superconductor Josephson
junctions were recently carried out, and modeled with single-level
quantum dots \cite{Park}.

Several issues related to averaging could be investigated in
the future:

First, it would be interesting to numerically average over a
distribution of the nonlocal Green's function for double 0D quantum
dots. A related issue is to implement quantum dots having dimension
that is large compared to the Fermi wave-length $\lambda_F$ instead of
the 0D quantum dots of the present paper. On the other hand,
{the ``metallic'' regime} of weak sample-to-sample
fluctuations of the quartet critical current can be addressed with
quasiclassics. For instance, discretized Usadel equations
{were used to evaluate multiple Andreev reflections}
in two-terminal devices in the dirty limit \cite{Cuevas-Usadel} and
three-terminal devices were addressed within assumptions about the
interface transparencies, also with Usadel equations
\cite{Cuevas-Pothier}.

Second, the use of Nazarov's circuit theory
implies that the Green's function is uniform within a node, which is
also satisfied by ballistic chaotic cavities \cite{Vanevic}. Thus
Nazarov's circuit theory is also appropriate to describe a ballistic
device if the coupling to the terminals is made through interfaces
with small cross-sections, to ensure that the metal behaves like a
chaotic cavity.

Third, an interesting perspective is to solve Eilenberger equations
for the four-terminal device shown in figure~\ref{fig:experiment}.

Fourth, it would also be interesting to average over voltage
fluctuations, the strength of which being controlled by the
electromagnetic environment.

To summarize this discussion, perspectives are about the interplay
between: (i) The ``Floquet effects'' that produce
``non-selfaveraging-like'' sharp resonance peaks in the
voltage-dependence of the quartet signal for single-channel contacts,
and (ii) The effect of averaging in space over extended contacts or in
energy over the voltage fluctuations induced by the electromagnetic
environment.

{Finally, probability current is conserved in the
  one-dimensional normal metal-superconductor junction treated by
  Blonder, Tinkham and Klapwijk \cite{BTK} [see Eqs.~(A7)-(A8) in this
    paper]. We leave as open the question of rigorously discussing
  probability conservation in the considered four-terminal device
  connected to normal leads, in connection with introducing the
  phenomenological Dynes parameter $\eta\ne 0$ and vanishingly small
  imaginary part of the local Green's function $B_I=0$ (as it was done
  here), or with $\eta=0$ and $B_I\ne 0$ (as it could be done in the
  future).}
  
{To conclude, the present paper suggests interest of
  ``quantum bath engineering'' multiterminal Josephson junctions in
  the circuits of cavity-quantum electrodynamics.}

\section*{Acknowledgements}

The author wishes to thank K. Huang, Y. Ronen and P. Kim for
stimulating discussions about their experiment. The author wishes to
thank R. Danneau for useful discussions and comments on the
manuscript, and F. Levy-Bertrand and her colleagues H. Cercellier,
K. Hasselbach, M.A. Measson for useful remarks during an informal
seminar on this topic. The author thanks the Infrastructure de Calcul
Intensif et de Donn\'ees (GRICAD) for use of the resources of the
M\'esocentre de Calcul Intensif de l’Universit\'e Grenoble-Alpes
(CIMENT). The author acknowledges support from the French National
Research Agency (ANR) in the framework of the Graphmon project
(ANR-19-CE47-0007).

\appendix

\section{{Spectrum in absence of coupling to the superconductors}}
\label{sec:evolution-t-0-app}

In this Appendix, we consider the ``phenomenological model of double
0D quantum dots with normal lead in series'', see
subsection~\ref{sec:remote-0D}, {and show that pairs
  of opposite energies emerge in the spectrum.}

{We start with the} simple limit where the four
superconducting leads are disconnected, i.e. $\Gamma=\Gamma'=0$, and
consider that the two quantum dots $D_x$ and $D_y$ are connected to
the tight-binding sites $N'_x$ and $N'_y$ by the tunneling amplitudes
$\Sigma^{(4)}_{N'_x,D_x} = \Sigma^{(4)}_{D_x,N'_x}$ and
$\Sigma^{(4)}_{N'_y,D_y} = \Sigma^{(4)}_{D_y,N'_y}$. The tight-binding
sites are self-connected by the Green's functions $g_{N'_x,N'_x}$ and
$g_{N'_y,N'_y}$, and connected to each other by the nonlocal Green's
functions $g_{N'_x,N'_y} = g_{N'_y,N'_x}$. We denote by
$g_{D_x,D_x}^A=g_{D_y,D_y}^A=1/(\omega-i\eta)$ the Green's functions
of the ``isolated'' quantum dots $D_x$ and $D_y$, see
Eqs.~(\ref{eq:H-dot-1})-(\ref{eq:H-dot-2}) where
$\epsilon_x=\epsilon_y=0$ in
Eqs.~(\ref{eq:H-x-0D})-(\ref{eq:H-y-0D}). In addition
$\tilde{g}_{D_x,D_x}$, $\tilde{g}_{D_y,D_y}$ are their counterparts
for the connected double 0D quantum dot, and by $\tilde{g}_{D_x,D_y}$
and $\tilde{g}_{D_y,D_x}$ are the corresponding nonlocal Green's
functions. The Dyson equations relate the $g$s to the $\tilde{g}$s
according to
\begin{eqnarray}
  \tilde{g}_{D_x,D_x}&=&g_{D_x,D_x}+
  g_{D_x,D_x} \Sigma^{(4)}_{D_x,N'_x} g_{N'_x,N'_x} \Sigma^{(4)}_{N'_x,D_x} \tilde{g}_{D_x,D_x}\\&&+
  \nonumber
  g_{D_x,D_x} \Sigma^{(4)}_{D_x,N'_x} g_{N'_x,N'_y} \Sigma^{(4)}_{N'_y,D_y} \tilde{g}_{D_y,D_x}\\
\tilde{g}_{D_y,D_x}&=&
  g_{D_y,D_y} \Sigma^{(4)}_{D_y,N'_y} g_{N'_y,N'_x} \Sigma^{(4)}_{N'_x,D_x} \tilde{g}_{D_x,D_x}\\&&+
  g_{D_y,D_y} \Sigma^{(4)}_{D_y,N'_y} g_{N'_y,N'_y} \Sigma^{(4)}_{N'_y,D_y} \tilde{g}_{D_y,D_x}
\nonumber
  ,
\end{eqnarray}
which leads to the secular equation
\begin{equation}
  \left|\begin{array}{cc}
  \omega-\Gamma_{D_x,N'_x,N'_x,D_x} & - \Gamma_{D_x,N'_x,N'_y,D_y}\\
  -\Gamma_{D_y,N'_y,N'_x,D_x} & \omega-\Gamma_{D_y,N'_y,N'_y,D_y}
  \end{array}\right|=0
  ,
\end{equation}
where $\Gamma_{D_x,N'_x,N'_x,D_x}=\Sigma^{(4)}_{D_x,N'_x} g_{N'_x,N'_x}
\Sigma^{(4)}_{N'_x,D_x}$, $\Gamma_{D_y,N'_y,N'_y,D_y}=\Sigma^{(4)}_{D_y,N'_y}
g_{N'_y,N'_y} \Sigma^{(4)}_{N'_y,D_y}$,
$\Gamma_{D_x,N'_x,N'_y,D_y}=\Sigma^{(4)}_{D_x,N'_x} g_{N'_x,N'_y}
\Sigma^{(4)}_{N'_y,D_y}$, $\Gamma_{D_y,N'_y,N'_x,D_x}=\Sigma^{(4)}_{D_y,N'_y}
g_{N'_y,N'_x} \Sigma^{(4)}_{N'_x,D_x}$. Assuming symmetric contacts
yields $\Gamma_{D_x,N'_x,N'_x,D_x} = \Gamma_{D_y,N'_y,N'_y,D_y} \equiv
\Gamma_{loc}$ and $ \Gamma_{D_x,N'_x,N'_y,D_y} =
\Gamma_{D_y,N'_y,N'_x,D_x} \equiv \Gamma_{non\,loc}$. The energy levels
are given by
\begin{eqnarray}
  \label{eq:ompm-1}
  \omega_{(\pm)} &=& \Gamma_{loc}\pm \Gamma_{non\,loc}\\
  &=& \left(\Sigma^{(4)}\right)^2 \left[B_R\pm A_R\right]
  + i \left(\Sigma^{(4)}\right)^2 \left[B_I\pm A_I\right]
  ,
  \label{eq:ompm-2}
\end{eqnarray}
{where $A_R$, $A_I$, $B_R$ and $B_I$ are given by
Eqs.~(\ref{eq:g-1})-(\ref{eq:g-2}).}

{Now, we assume $B_R=B_I=0$, as in
  section~\ref{sec:num-2nodes}. Then,
  Eqs.~(\ref{eq:ompm-1})-(\ref{eq:ompm-2}) yield energy levels
  $\omega_{(\pm)}=\pm\Sigma^{(4)}\left(A_R+iA_I\right)$ having
  opposite real and imaginary parts.}

\section{$V=0^+$ adiabatic limit}
\label{app:0+}
In this Appendix, we examine the $V=0^+$ adiabatic limit of
four-terminal Josephson junctions containing a single or two quantum
dots (see subsections~\ref{sec:0D} and~\ref{sec:ex-double} below).

\subsection{Single quantum dot}
\label{sec:0D}

We start with single 0D quantum dots in the $V=0^+$ adiabatic limit,
summarizing a fraction of the Supplemental Material of our previous
paper~II \cite{paperII}.

The Dyson equations take the following form for the considered 0D
quantum dot connected to $p_0$ superconducting leads by the tunnel
amplitudes $\hat{\Sigma}^{(5)}_{D_x,S_{n_p}} =
\hat{\Sigma}^{(5)}_{S_{n_p},D_x}$, with $n_p=1,...,p_0$:
\begin{equation}
  \label{eq:Gxx}
  \hat{G}_{D_x,D_x}=\hat{g}_{D_x,D_x}+\hat{g}_{D_x,D_x} \sum_{n_p=1}^{p_0}
  \hat{\Sigma}^{(5)}_{D_x,S_{n_p}} \hat{g}_{S_{n_p},S_{n_p}}
  \hat{\Sigma}^{(5)}_{S_{n_p},D_x} \hat{G}_{D_x,D_x}
  .
\end{equation}
In the infinite-gap limit, Eq.~(\ref{eq:Gxx}) can be expressed with
the infinite-gap Hamiltonian $\hat{\cal
  H}_{eff,\,single\,dot}^{(\infty)}$:
\begin{equation}
  \hat{G}_{D_x,D_x}^A=\left(\omega-i\eta-\hat{\cal H}_{eff,\,single\,dot}^{(\infty)}\right)^{-1}
  ,
\end{equation}
where
\begin{equation}
  \hat{\cal H}_{eff,\,single\,dot}^{(\infty)}=\sum_{n_p=1}^{p_0}
  \hat{\Sigma}^{(5)}_{D_x,S_{n_p}} \hat{g}_{S_{n_p},S_{n_p}}
  \hat{\Sigma}^{(5)}_{S_{n_p},D_x} .
\end{equation}
Specifically, we obtain the following with $p_0=4$ superconducting
leads:
\begin{equation}
  \hat{\cal H}_{eff,\,single\,dot}^{(\infty)}
  =\left(\begin{array}{cc} 0 &\gamma_{D_x,D_x}\\\left(\gamma_{D_x,D_x}\right)^* & 0
  \end{array}\right)
  ,
\end{equation}
where
\begin{eqnarray}
  \gamma_{D_x,D_x}&=&\Gamma_a \exp(i\varphi_a)+
  \Gamma_b \exp(i\varphi_b)\\&+&
  \Gamma_{c,1} \exp(i\varphi_{c,1})+
  \Gamma_{c,2} \exp(i\varphi_{c,2})
  ,\nonumber
\end{eqnarray}
and $\Gamma_{n_p}=\left(\Sigma_{D_x,S_{n_p}}^{(5)}\right)^2/W$
parameterizes the line-width broadening of the quantum dot level in
the normal state.

The $(S_{c,1},S_{c,2})$ superconducting leads can be gathered into the
single $S_{c,eff}$ coupled by
\begin{equation}
  \label{eq:Gamma-eff-0D}
  \Gamma_{c,eff}=\Gamma_{c,1} \exp\left(i\varphi_{c,1}\right)
  + \Gamma_{c,2} \exp\left(i\varphi_{c,2}\right)
  .
\end{equation}

Using the identical $\Gamma_{c,1}=\Gamma_{c,2}\equiv \Gamma_c$ and the
gauge given by Eqs.~(\ref{eq:gauge1})-(\ref{eq:gauge2}) yields
\begin{equation}
  \Gamma_{c,eff}=\Gamma_c \left[1+\exp\left(i\Phi\right)\right]
  \exp\left(i\varphi_{c,1}\right)
  .
  \label{eq:c-eff}
\end{equation}

It was shown in the Supplemental Material of our previous paper~II
\cite{paperII} that nonsymmetric coupling to the superconducting leads
can produce inversion between $\Phi=0$ and $\Phi=\pi$ in the
infinite-gap limit. But Eq.~(\ref{eq:c-eff}) implies
$|\Gamma_{c,eff}|=2\Gamma_c$ if $\Phi=0$ and $\Gamma_{c,eff}=0$ if
$\Phi=\pi$ for symmetric couplings to the superconducting leads, {\it i.e.} the
quartet current at $\Phi=\pi$ is vanishingly small, thus it is
automatically smaller than at $\Phi=0$.

\begin{widetext}
\subsection{Double 0D quantum dot}
\label{sec:ex-double}
In this subsection, we provide simple argument for emergence of
inversion in the double 0D quantum dot with normal lead in parallel in
figure~\ref{fig:geometrie-4T-double-dot}, in the limits
$eV/\Delta=0^+$ and $\Gamma'=0$.

In the infinite-gap limit, the $4\times 4$ Hamiltonian of a double 0D
quantum dot with normal leads in parallel is given by
\begin{equation}
  \hat{\cal H}_{eff,\,double\,dot}^{(\infty)}=\left(\begin{array}{cccc}
    0 & \gamma_{D_x,D_x} & \Sigma^{(0)} & 0\\
    \left(\gamma_{D_x,D_x}\right)^* & 0 & 0 & - \Sigma^{(0)}\\
    \Sigma^{(0)} & 0 & 0 & \gamma_{D_y,D_y}\\
    0 & -\Sigma^{(0)} & \left(\gamma_{D_y,D_y}\right)^* & 0
  \end{array}\right)
  ,
  \label{eq:H-infinite-4x4}
\end{equation}
{where we assumed $\Gamma'=0$, and $\gamma_{D_x,D_x}$,
  $\gamma_{D_y,D_y}$ are given by the above
  Eqs.~(\ref{eq:gamma-Dx-Dx})-(\ref{eq:gamma-Dy-Dy}).} Squaring the
infinite-gap Hamiltonian given by Eq.~(\ref{eq:H-infinite-4x4}) leads
to
\begin{eqnarray}
  && \left(\hat{\cal H}_{eff,\,double\,dot}^{(\infty)}\right)^2=\\&&
  \nonumber
  \left(\begin{array}{cccc}
    \left|\gamma_{D_x,D_x}\right|^2+\left(\Sigma^{(0)}\right)^2 & 0 & 0 &
    - \Sigma^{(0)}\left[\gamma_{D_x,D_x}-\gamma_{D_y,D_y}\right]\\
    0 & \left|\gamma_{D_x,D_x}\right|^2+\left(\Sigma^{(0)}\right)^2
    & \Sigma^{(0)}\left[\left(\gamma_{D_x,D_x}\right)^*
      -\left(\gamma_{D_y,D_y}\right)^*\right] & 0\\
    0 & \Sigma^{(0)}\left[\gamma_{D_x,D_x}-\gamma_{D_y,D_y}\right] &
    \left|\gamma_{D_y,D_y}\right|^2+\left(\Sigma^{(0)}\right)^2 & 0\\
    -\Sigma^{(0)}\left[\left(\gamma_{D_x,D_x}\right)^*
      -\left(\gamma_{D_y,D_y}\right)^*\right] & 0 & 0
    & \left|\gamma_{D_y,D_y}\right|^2+\left(\Sigma^{(0)}\right)^2
  \end{array}\right)
  ,
\end{eqnarray}
which decouples into the following $2\times 2$ blocks:
\begin{eqnarray}
  \label{eq:HH1}
  \left[\hat{\cal H}^2\right]^{(0)}_{2\times 2} =
  \left( \begin{array}{cc}
    \left|\gamma_{D_x,D_x}\right|^2+\left(\Sigma^{(0)}\right)^2 & - \Sigma^{(0)}\left[\gamma_{D_x,D_x}-\gamma_{D_y,D_y}\right]\\
    -\Sigma^{(0)}\left[\left(\gamma_{D_x,D_x}\right)^*
      -\left(\gamma_{D_y,D_y}\right)^*\right] & 
    \left|\gamma_{D_y,D_y}\right|^2+\left(\Sigma^{(0)}\right)^2
  \end{array}\right)
\end{eqnarray}
and
\begin{eqnarray}
    \left[\hat{\cal H}^2\right]^{(2)}_{2\times 2}=
  \left( \begin{array}{cc}
\left|\gamma_{D_x,D_x}\right|^2+\left(\Sigma^{(0)}\right)^2
    & \Sigma^{(0)}\left[\left(\gamma_{D_x,D_x}\right)^*
  -\left(\gamma_{D_y,D_y}\right)^*\right] \\
\Sigma^{(0)}\left[\gamma_{D_x,D_x}-\gamma_{D_y,D_y}\right] &
\left|\gamma_{D_y,D_y}\right|^2+\left(\Sigma^{(0)}\right)^2
  \end{array} \right)
  .
\end{eqnarray}
\end{widetext}
Thus,
\begin{eqnarray}
\gamma_{D_x,D_x}-\gamma_{D_y,D_y}&=&\Gamma_a \exp\left(i\varphi_a\right)
  + \Gamma_{c,1} \exp\left(i\varphi_{c,1}\right)\\&-& \Gamma_b \exp\left(i\varphi_b\right)
  - \Gamma_{c,2} \exp\left(i\varphi_{c,2}\right)
  \nonumber
\end{eqnarray}
and the coupling to the effective $S_{c,eff}$ is now given by the difference
\begin{equation}
  \label{eq:Gamma-eff-double}
  \Gamma_{c,eff}= \Gamma_{c,1} \exp\left(i\varphi_{c,1}\right)
  - \Gamma_{c,2} \exp\left(i\varphi_{c,2}\right)
\end{equation}
instead of the previous Eq.~(\ref{eq:Gamma-eff-0D}) for a single 0D
quantum dot. Eq.~(\ref{eq:Gamma-eff-double}) goes to
\begin{equation}
  \Gamma_{c,eff}(\Phi)= \Gamma \left[1-\exp\left(i\Phi\right)\right] \exp\left(i\varphi_{c,1}\right)
\end{equation}
in the {considered} limit $\Gamma_{c,1}=\Gamma_{c,2} \equiv \Gamma$ of
symmetric couplings. Thus, {the interference}
$|\Gamma_{c,eff}|(\Phi=0)=0$ and $|\Gamma_{c,eff}|(\Phi=\pi)=2\Gamma$
yields inversion between $\Phi=0$ and $\Phi=\pi$ with symmetric
coupling to the superconducting leads. This contrasts with absence of
inversion for single 0D quantum dots, see section~\ref{sec:0D} in this
Appendix.

\end{document}